 \documentclass[12pt,preprint]{aastex}

\newcommand{\kms}{km\thinspace s$^{-1}$}

\begin{document}

\title{A CO Survey of Young Planetary Nebulae\altaffilmark{1} }


\author{P. J. Huggins\altaffilmark{2},  
R. Bachiller\altaffilmark{3}, P. Planesas\altaffilmark{3},
T. Forveille\altaffilmark{4, 5}, P. Cox\altaffilmark{6}
 } 

\altaffiltext{1}{Based on observations
carried out with the IRAM 30 m telescope. IRAM is
supported by INSU/CNRS (France), MPG (Germany), and IGN (Spain).}
\altaffiltext{2}{Physics Department, New York University, 4 Washington
Place, New York NY 10003}
\altaffiltext{3}{IGN Observatorio Astron\'omico Nacional, Apartado
112, E-28003 Alcal\'a de Henares, Spain}
\altaffiltext{4}{Observatoire de Grenoble, B.P. 53X, 38041 Grenoble
Cedex, France}
\altaffiltext{5}{Canada-France-Hawaii Telescope, PO Box 1597, Kamuela, 
HI 96743}  
\altaffiltext{6}{Institut d'Astrophysique Spatiale, Universit\'e de
Paris Sud, 91405 Orsay, France}



\begin{abstract}
We report the results of a sensitive survey of young planetary nebulae
in the CO $J=2-1$ line that significantly increases the available data
on warm, dense, molecular gas in the early phases of planetary nebula
formation. The observations were made using the IRAM 30~m telescope
with the 3$\times$3 pixel Heterodyne Receiver Array (HERA). The array
provides an effective means of discriminating the CO emission of
planetary nebulae in the galactic plane from contaminating emission of
interstellar clouds along the line of sight. 110 planetary nebulae
were observed in the survey and 40 were detected. The results increase
the number of young planetary nebulae with known CO emission by
approximately a factor of two. The CO spectra yield radial velocities
for the detected nebulae, about half of which have uncertain or no
velocity measurements at optical wavelengths. The CO profiles range
from parabolic to double-peaked, tracing the evolution of structure in
the molecular gas. The line widths are significantly larger than on
the Asymptotic Giant Branch, and many of the lines show extended
wings, which probably result from the effects on the envelopes of high
velocity jets.

\end{abstract}


\section{Introduction}

Observations of molecular gas have played a key role in developing our
current understanding of stellar evolution from the Asymptotic Giant
Branch (AGB) through the planetary nebula (PN) phase. Stars evolving
on the upper AGB undergo an increasing rate of mass loss which forms a
dense, slowly expanding circumstellar envelope of dust and molecular
gas.  The mass loss eventually terminates the evolution on the AGB,
and when the surface temperature of the star becomes high enough to
emit a significant flux of ultraviolet radiation ($T\ga 30,000$~K), it
begins to photo-ionize the circumstellar material and form an
optically visible PN.

Observations of neutral gas around bona fide PNe have been crucial in
the development of this general scenario, and numerous atomic and
molecular species have been observed. These include H\,{\sc i},
C\,{\sc i}, O\,{\sc i}, and other neutral atoms (e.g., Rodriguez,
Goss, \& Williams 2002; Bachiller et al. 1994; Liu et al. 2001;
Dinerstein, Sneden, \& Uglum 1995), as well as H$_2$ (e.g., Kastner et
al. 1996), OH (e.g., Zijlstra et al. 2001), CO (e.g., Huggins et
al. 1996), and other, rarer molecular species such as HCO$^+$ and CN
(e.g., Bachiller et al. 1997). Millimeter CO emission has proved to be
the most generally useful probe of the warm molecular component of the
neutral gas for the same reasons that it is a standard probe of the
precursor AGB envelopes: CO is relatively abundant and the millimeter
transitions are easily excited. It can therefore be used to study the
mass, structure, and kinematics of the bulk of the molecular gas.

Most of the known CO envelopes around PNe were discovered in dedicated
CO surveys carried out by Huggins \& Healy (1989) using the NRAO 12~m
telescope, and by Huggins et al. (1996) using the SEST 15~m and IRAM
30~m telescopes. The compilation in Table~3 of Huggins et al. (1996)
lists 44 PNe detected in CO. Several of these have provided examples
for detailed study at high angular resolution, and interpretation of
the ensemble of the CO observations forms an integral part of the
evidence for the evolutionary scenario outlined above.

Several recent lines of investigation, including optical imaging of PNe
(e.g., Sahai \& Trauger 1998; Lopez 2003), kinematic studies of proto-PNe
(e.g., Bujarrabal et al. 2001), and observations of individual
molecular envelopes (e.g., He 3-1475, Huggins et al. 2004; AFGL 618,
Cox et al. 2003), have focused attention on the complex geometry of PN
formation. The underlying mechanisms are not well understood, but they
likely involve effects that influence the survival of, or
affect the characteristics of, the molecular gas found in young PNe.
Accordingly we have undertaken a deeper search for CO, especially in
young PNe, to significantly extend the sample where the
characteristics of the molecular gas are known.
 
Based on considerations of line opacity, atmospheric transparency,
telescope beam size, and receiver sensitivity, the 230~GHz CO $J=2-1$
line is the optimal transition for CO searches of PN envelopes, where the
emission is distributed on small angular scales and is optically thin,
or partially thin, in most cases.  In the time since our previous
survey work, an important increase in capability at 230~GHz has been
the development of multiple beam receivers. PNe are distributed close to
the galactic plane, and a sensitive search in CO is inevitably
contaminated by interstellar CO emission. The use of a multiple beam
array on the IRAM 30~m telescope for the present survey observations
proves to be a very efficient way to address this issue.

The plan of the rest of this paper is as follows. Section 2 describes
the sample selection and the observations; Sections 3 and 4 present
the results of the survey, in the form of a table, spectra, and
comments on individual objects; and Section 5 gives gives a brief
discussion.  A detailed analysis of the results will given in a
separate paper.

\section{The Survey}

\subsection{Survey Sample}

The survey PNe were taken from the Strasbourg-ESO Catalog of Galactic
PNe (Acker et al. 1992) and were chosen to include a major component
of relatively young PNe with a broad range of spectroscopic
characteristics.  The PNe were selected to have small angular
diameters ($\la$ the telescope beam size whose FWHM is $12\arcsec$),
relatively high fluxes in H$\alpha$ (with a few exceptions
$>10^{-13}$~erg~cm$^{-2}$~s$^{-1}$, uncorrected for reddening), and
consequently moderate to high surface brightness in the Balmer lines
and the radio continuum. The PNe lie close to the galactic plane, and
the practical declination limit set by the telescope location is $\sim
-$25\arcdeg.  In addition, the PNe were distributed among different
spectroscopic groups, based on line ratios given in the PN catalog, or
updated from the literature: 1) [O\,{\sc iii}] $\lambda$5007/H$\beta <
6.0$ ; 2) [O\,{\sc iii}] $\lambda$5007/H$\beta > 6.0$ and [N\,{\sc
ii}] $\lambda$ 6584/H$\alpha > 1.0$; 3) [O\,{\sc iii}]
$\lambda$5007/H$\beta > 6.0$ and [N\,{\sc ii}] $\lambda$ 6584/H$\alpha
< 1.0$. These line ratios are not very sensitive to the reddening, but
are subject to uncertainties related to the slit position and size
relative to the nebula. A fourth group of miscellaneous PNe (see
below) was also observed as time permitted in the program.

The angular sizes of relatively young PNe are expected to be quite
small at distances outside our local neighborhood $\sim 1$~kpc, which
has been explored in CO by earlier observations. For example, the well
known PN NGC~6720 (the Ring Nebula) which is used as an intermediate
benchmark below, has an angular diameter of $76\arcsec$ at a distance
of 450~pc (O'Dell et al. 2002), and would have an angular diameter of
$\la 11\arcsec$ at distances $\ga 3$~kpc. More compact PNe would have
even smaller angular diameters, and such objects are well suited to
line searches made with single observations centered on the nebula.
The average surface brightness of the PNe in the Balmer lines or the
radio continuum is independent of distance, and is a useful guide to
the degree of evolution of the nebulae. For discussion here we use the
surface brightness $T_{\rm B}$ at a frequency of 1.4~GHz, using the
fluxes measured directly at 1.4 GHz from Condon \& Kaplan (1998) or
the equivalent from the (de-reddened) Balmer lines, assuming optically
thin emission. The emission at 1.4~GHz becomes optically thick before
that at higher radio frequencies, but high opacity affects relatively
few of the PNe and this does not substantially affect our overall
discussion.  The distribution of $T_{\rm B}$ for the whole PN sample
ranges from 4~K to $\ga 10,000$~K, with 1st, 2nd and 3rd quartiles at
at 110~K, 800~K, and 4,200~K, respectively.

It is well known that the surface brightness of PNe roughly correlates
with nebula size (e.g., Phillips 2002, and references therein), and is
therefore related to the dynamical age of the nebulae. We do not
discuss this in great detail but note as benchmarks that $T_{\rm
B}=3,500$~K for the compact PN BD+30\arcdeg3639 and $T_{\rm B}= 65$~K
for the more evolved PN NGC~6720, and their expansion ages based on
robust proper motions measured with HST are 800~yr and 1,500~yr,
respectively (Li et al. 2002; O'Dell et al. 2002).  32\% of the survey
PNe have $T_{\rm B} \ga 3,500$~K and 84\% have $T_{\rm B}\ga
65$~K. Thus the bulk of the observed PNe, with the exception of those
with the lowest values of $T_{\rm B}$, are expected to be younger (and
in many cases much younger) than $\sim 2000$ yr. Our selection
criteria strongly discriminate against large, evolved, low surface
brightness PNe. Some of these are known to retain significant CO
envelopes, as exemplified by the Helix Nebula (Young et al. 1999),
$T_{\rm B} \sim 1$~K, age $\sim$ 10$^4$~yr, but the objective of the
present work is to study the earlier evolution of the envelopes.

It is known from our previous observations (Huggins et al. 1996) that
CO envelopes of PNe show large differences which are likely related to
the mass of the central stars, and possibly additional factors as
well. The distribution of the survey PNe over different spectroscopic
groups is intended to provide information to investigate this further.
Of course the relation between age and nebula excitation is a complex
one: young PNe can have a range of stellar temperatures because the
speed of the temperature increase of the star depends strongly on the
stellar mass (e.g., Gesicki \& Zijlstra 2000). The nebula excitation 
is therefore not a direct measure of the nebula age, but alone, or in
combination with the brightness temperature may cast light on, or
serve as a predictor of the CO emission.

Based on the [O\,{\sc iii}] $\lambda$5007/H$\beta$ line ratio the
first group (40 objects) selects low excitation PNe (excitation
classes $\la 2$ on the scale of Webster 1975) with central star
temperatures $\la 45,000$~K (e.g. Kaler \& Jacoby 1991).  The second
group (31 objects) selects higher excitation PNe with strongly
enhanced [\ion{N}{2}] emission and is likely to include the higher
mass progenitor stars which pass rapidly through the low excitation
phase; the third group (28 objects) acts as a control.  A small,
additional fourth group (11 objects) of miscellaneous PNe with
incomplete spectroscopy, or with larger angular diameters and
typically in the lowest quartile of $T_{\rm B}$, was also observed as
time in the program allowed.

\subsection{Observations}

The survey observations were made in the CO $J=2-1$ line at
230.538~GHz using the IRAM 30~m telescope at Pico Veleta, Spain,
in March 2002 and May 2003.

The receiver system used was the multibeam Heterodyne Receiver Array
(HERA), which consists of nine receivers arranged in a regular
$3\times3$ grid with a spacing on the sky of 24\arcsec. The beam size
of each ``pixel'' at 230~GHz is 12\arcsec\ (FWHM). For a gaussian
source of size 12\arcsec\ (FWHM) centered in the array, a negligible
fraction ($<1\%$) of the signal in the central pixel is recorded in
the outer pixels. The backend used was the VESPA autocorrelator, which
was configured to provide $225 \times 1.63$~\kms\ channels for each of
the nine receivers during the 2002 observing run, and $409 \times
1.63$~\kms\ channels during the 2003 observing run. A wide spectral
coverage is essential because the CO lines of PN envelopes often have
broad wings and the systemic radial velocities of many of the PNe have
not previously been measured at optical wavelengths (see Section~5).

The telescope pointing was regularly monitored during the observations
using reference sources, and is generally expected to be accurate
within $\sim 2\arcsec$, although it is somewhat more uncertain during
periods of high wind or rapidly changing weather conditions. The
observations were made by switching in azimuth by 4\arcmin\ using the
wobbler. The spectra were calibrated using the chopper wheel
technique, and the intensities are reported here as main beam
temperatures, using a beam efficiency of 0.51.  The velocity scale
used for the spectra is heliocentric.

The coordinates of the PNe were obtained from a compilation of optical
positions (Acker et al. 1992) and accurate radio positions measured
with the VLA (Zijlstra, Pottasch, \& Bignell 1989; Aaquist \& Kwok
1990; Condon \& Kaplan 1998). In almost all cases the coordinates are
known to within $\sim 1\arcsec$, and in the least favorable cases to
within a few arc seconds. The coordinates are therefore not a
significant source of uncertainty for the observations. Nearly all the
PNe lie at low galactic latitude ($b$), with a mean value of $|b|$ of
$3\fdg 8$.

Each PN was observed centered in the array. In all cases where the
envelope was detected there was no significant envelope emission in
the surrounding pixels.  Interstellar (IS) lines were seen along most
lines of sight, and they appear as emission (if they occur in the on
position) or as dips (if they occur in the off position). In some
cases the lines are narrow, isolated features and there is little
confusion with the PN emission. In other cases, where there are
multiple components which may be merged, comparison of all the spectra
from the array is very efficient in discriminating the IS from the PN
components.  The spectra presented in the next section have been
obtained by subtracting the average of the outer pixels from the
central pixel. This has little effect on the S/N, and is helpful in
reducing the strength of the IS components, especially in cases where
they are fairly uniform or have a smooth gradient across the field of
the array.

\section{Results}
 
110 PNe were observed in the survey and 40 were detected.  The results
of the observations are summarized in Table~1. Column (1) gives the PN
name in galactic notation (ordered by galactic longitude); column (2)
the common name; column (3) the rms noise level of the observations;
columns (4)--(7) the parameters of the CO emission in the cases where the
PN is detected (see below); column (8) the heliocentric velocity interval over
which there is significant interstellar contamination; and column (9)
comments.
  
The spectra of the PNe in which CO emission is detected or tentatively
detected are presented in Fig.~1.  It can be seen that the spectra
exhibit a large range in S/N ratio and a variety of line profiles. The
parameters of the emission given in Table~1 refer to the integrated
line intensity ($I$), the heliocentric systemic velocity in CO
($V_{\rm o}$), the line width at zero intensity ($\Delta V$), and the
peak line temperature ($T_{\rm p}$).  In the spectra where the S/N
ratio is sufficiently high that the profile shape can be discerned,
$I$ is obtained by direct integration of the profile, $V_{\rm o}$ is
estimated where possible from the symmetry of the line (e.g., the
average velocity of the peaks of a double-peaked profile), and $\Delta
V$ and $T_{\rm p}$ are determined by inspection; because the intrinsic
profiles are unknown, larger values of $\Delta V$ from broad wings at
or below the noise level of the spectra cannot be ruled out. In the
spectra with low S/N where the profile shape cannot be discerned, the
parameters are obtained by a Gaussian fit to the line; the line width
given in Table~1 is the FWHM, and it is given as a lower limit to
$\Delta V$.

\section{Comments on Individual Spectra}

Here we comment on the individual spectra in Fig.~1, point out IS
features that appear in the profiles, and compare with available
optical radial velocities ($V_{\rm opt}$). These are taken from the
compilation by Durand, Acker, \& Zijlstra (1998), except where
explicitly cited. Previous CO observations at other telescopes are
also noted.

002.6+04.2 (Th 3-27) Tentative 6-$\sigma$ detection.  $V_{\rm o}$ consistent
    with $V_{\rm opt}$ ($-135.3\pm10.1$~\kms). The narrow CO spike at $V =
    -13$~\kms\ is IS.

003.8+05.3 (H 2-15) Tentative 5-$\sigma$ detection.  $V_{\rm o}$ consistent
    with $V_{\rm opt}$ ($-63.2\pm3.7$~\kms).

008.3$-$01.1 (M 1-40) Roughly flat-topped profile with severe IS contamination;
    the emission at $V \ga 0$~\kms\ is probably all IS.  $V_{\rm o}$
    consistent with $V_{\rm opt}$ ($-34.4\pm2.0$~\kms).

008.6$-$07.0 (He 2-406) Double-peaked profile. $V_{\rm o}$ consistent with
    $V_{\rm opt}$ ($+28.2\pm5.2$~\kms).

010.1+00.7 (NGC 6537) Strong, triangular profile, with IS
contamination on the red wing. $V_{\rm o}$ of peak differs from
$V_{\rm opt}$ ($-16.9\pm1.9$~\kms) by 13~\kms, suggesting some
structural asymmetry. In deep images the nebula is much more extended
than the $10\arcsec$ listed in the Strasbourg-ESO catalog, but CO is
seen only in the central pixel, and likely arises in the dense inner
torus whose diameter in HST images is $\sim 5\arcsec$ (Huggins \&
Manley 2005). A broad CO(1--0) feature at $-$8~\kms\ (+5~\kms\ LSR)
reported by Zhang et al. (2000) is significantly different from the
spectrum reported here and is likely at least partly interstellar.

011.9+04.2 (M 1-32) Tentative 3-$\sigma$ detection.  $V_{\rm o}$ consistent
    with $V_{\rm opt}$ ($-90.6\pm7.6$~\kms).

016.0$-$04.3 (M 1-54) $V_{\rm o}$ of the CO peak is offset by $\sim 15$~\kms\
    from $V_{\rm opt}$ ($-47.4\pm4.2$~\kms), so the tentative, second CO
    feature at $V = -87$~\kms\ may be part of a broad,
    double-peaked profile.

019.9+00.9 (M 3-53)  Probably a flat-topped profile, but strong IS (spike/dip)
    contamination of the blue wing. $V_{\rm o}$ consistent with
    uncertain $V_{\rm opt}$ ($+20.6\pm14.0$~\kms).

021.1$-$05.9 (M 1-63) Double-peaked profile. $V_{\rm o}$ consistent with
    $V_{\rm opt}$ ($+8.5\pm10.3$~\kms).

021.7$-$00.6 (M 3-55) Strong, parabolic profile, with severe IS
    (spike/dip) contamination; emission at $V \ga 40$~\kms\ is
    probably all IS. The broad wing to the blue appears to belong to
    the PN.  $V_{\rm o}$ consistent with uncertain $V_{\rm opt}$
    ($+9.7\pm15.0$~\kms).

021.8$-$00.4 (M 3-28) Strong, double peaked profile with IS (spike)
    contamination on red wing and at higher velocities.  A CO spectrum
    obtained with the SEST 15~m telescope was previously reported by
    Gomez, Rodriguez, \& Garay (1992).  $V_{\rm o}$ consistent with
    uncertain $V_{\rm opt}$ ($+4.7\pm15.0$~\kms).

023.9$-$02.3 (M 1-59) Single component with prominent broad wings. A
    CO spectrum obtained with the JCMT 15~m telescope was previously
    reported by Gussie \& Taylor (1995). Emission spike at $V =
    -8$~\kms\ is IS. $V_{\rm o}$ consistent with uncertain $V_{\rm
    opt}$ ($+85.9\pm14.0$~\kms).

025.9$-$00.9 (Pe 1-14) Single component profile with series of narrow IS
    spike/dip features.  $V_{\rm opt}$ is unknown.

025.9$-$10.9 (Na 2) Double-peaked profile. $V_{\rm o}$ consistent with
    uncertain $V_{\rm opt}$ ($+97.8\pm12.0$~\kms).

031.7+01.7 (PC 20) Single-peaked profile.  $V_{\rm opt}$ is unknown.

032.7+05.6 (K 3-4) Single peaked profile with broad wing on red side; IS
    spike/dip contamination on the blue wing from 0 to $-$18~\kms.
    $V_{\rm opt}$ is unknown.

039.8+02.1 (K 3-17) Strong, remarkably broad, roughly flat-topped profile,
    possibly with additional weak wings; the dips in the profile at $V
    = 13$--18~\kms\ are IS.  $V_{\rm opt}$ is unknown.

043.0$-$03.0 (M4-14) Strong, asymmetric double-peaked profile, with broad
    wings.  $V_{\rm o}$ consistent with uncertain $V_{\rm opt}$
    ($+31.7\pm15.0$~\kms).

047.1+04.1 (K 3-21) Tentative 5-$\sigma$ detection.  $V_{\rm opt}$ is unknown.

055.3+02.7 (He 1-1) Tentative 5-$\sigma$ detection. $V_{\rm o}$
differs by $\sim 42$~\kms\ from $V_{\rm opt} \sim -35$~\kms\ from
Guerrero et al. (1999), but roughly coincides with one of the main
optical components.

055.6+02.1 (He 1-2) Flat-topped or concave profile with IS spike on the blue
    wing at $V = -8$~\kms.  $V_{\rm opt}$ is unknown.

056.0+02.0 (K 3-35) Flat-topped profile with IS spike on the blue wing
at $V = -8$~\kms.  A CO spectrum obtained with the NRAO 12~m telescope
was previously reported by Dayal \& Bieging (1996). $V_{\rm o}$ agrees
with H$_2$O maser velocities in the core region (Miranda et al. 2001).
$V_{\rm opt}$ ($-9\pm2$~\kms, converted from LSR from Miranda et
al. 2000) is offset by $\sim 12$~\kms.

060.1$-$07.7 (NGC 6886) $V_{\rm o}$ consistent with very
    uncertain $V_{\rm opt}$ ($-35.8\pm21.1$~\kms).

060.5$-$00.3 (K 3-45) Strong, double-peaked profile; narrow dip at $V =
   +15$~\kms\ and spike at $V = -32$~\kms\ are IS.  $V_{\rm opt}$ is
   unknown.
 
062.4$-$00.2 (M 2-48) Strong, (probably) double-peaked profile, with
IS contamination of red side; narrow dip at $V = -11$~\kms\ is also
IS. $V_{\rm o}$ consistent with $V_{\rm opt} \sim -15$~\kms\ in core
from Lopez-Martin et al. (2002). 
 
067.9$-$00.2 (K 3-52) Double or single component with broad wings affected by
    strong IS contamination: dips in profile at $V = -17$, $-$4 and
    +4~\kms\ are IS. $V_{\rm opt}$ is unknown.

068.3$-$02.7 (He 2-459) Roughly parabolic profile with IS dips just
outside the red wing.  The very uncertain $V_{\rm opt}$
($-72.0\pm20.0$~\kms) differs from $V_{\rm o}$ by $\sim
33$~\kms. CO(1--0) features at $-$9 and $-$15~\kms\ (+9 and +3~\kms\ LSR)
reported by Zhang et al. (2000) are likely interstellar.

069.6$-$03.9 (K 3-58) Strong, double-peaked profile. $V_{\rm opt}$ is unknown.

074.5+02.1 (NGC 6881) Probably double peaked profile, but strong IS contamination
    (spikes) on blue side. $V_{\rm o}$ consistent with $V_{\rm opt}$
    ($-14.4\pm2.3$~\kms).

079.6+05.8 (M 4-17) Flat-topped or concave profile. $V_{\rm o}$ consistent with
    very uncertain $V_{\rm opt}$ ($-26.0\pm40.0$~\kms).

091.6$-$04.8 (K 3-84) Flat-topped profile.  $V_{\rm opt}$ is unknown.

094.5$-$00.8 (K 3-83) Asymmetric or double-peaked profile, affected by IS
    contamination on red side; dip at $V = -57 $~\kms\ and just beyond
    the red wing are IS. $V_{\rm opt}$ is unknown.

104.4$-$01.6 (M 2-53) Steep-sided, double-peaked profile.  Deep IS dip at $V =
-43 $~\kms\ affects the extreme red wing of the profile. $V_{\rm o}$
consistent with very uncertain $V_{\rm opt}$ ($-62.0\pm40.0$~\kms).

107.6$-$13.3 (Vy 2-3) Tentative 5$\sigma$ detection.  Line is narrow for a PN
    but it could be the peak of a broader unseen component. It could
    be IS, but it is not seen in any of the surrounding 8 pixels, and
    $V_{\rm o}$ consistent with well determined $V_{\rm opt}$
    ($-49.5\pm3.8$~\kms).

119.6$-$06.7 (Hu 1-1) Single-peaked, asymmetric profile.  The well determined
    $V_{\rm opt}$ ($-53.7\pm3.0$) is offset by $\sim 14$~\kms\ from
    the $V_{\rm o}$ peak, in the direction of the extended wing.

130.4+03.1  (K 3-92) Double-peaked profile.  $V_{\rm o}$ consistent 
with $V_{\rm opt}$ ($-61.7\pm2.8$~\kms).

149.0+04.4 (K 4-47) Strong, roughly parabolic profile. $V_{\rm opt}$
consistent with $V_{\rm opt} \sim -37$~\kms\ from Corradi et al. (2000).
 
153.7$-$01.4  (K 3-65) Double-peaked profile; spike at $V = -26 $~\kms\ is
    IS. $V_{\rm opt}$ is unknown.

184.6+00.6 (K 3-70) Flat-topped profile. $V_{\rm o}$ consistent with $V_{\rm
    opt}$ ($+26.9\pm4.5$~\kms).

359.8+06.9 (M 3-37) Strong, double-peaked profile; spike at $V = -5 $~\kms\ is
    IS. $V_{\rm o}$ consistent with $V_{\rm opt}$ ($-74.2\pm10.2$~\kms).

\section{Discussion}

\subsection{Detection Rates}
  
The observations reported here significantly increase the available
data on the CO envelopes of PNe.  The compilation by Huggins et
al. (1996) lists 44 PNe detected in CO, and only a handful of other
PNe have been detected in searches in the intervening period (e.g.,
Josselin \& Bachiller 2000).  One focus of the earlier work was large,
highly evolved PNe with low surface brightness. The new sample has a
higher proportion (85\%) of PNe with moderate to high surface
brightness ($T_{\rm B} \ga 65$~K using the benchmark of Section 2.1),
compared to 64\% of the earlier compilation, and roughly doubles the
number of relatively young PNe with known CO envelopes. The 70
sensitive upper limits reported here also improve and extend the
earlier data on PNe where the molecular component has been destroyed.
 
The overall detection rate of the new sample is 36\%, and the rate
shows no significant variation with $T_{\rm B}$. On the other hand,
the detection rates for the spectroscopic groups 1--4 are highly
non-uniform at 15\%, 71\%, 36\%, and 18\%, respectively; this
variation is significantly different from that expected on the
hypothesis of a uniform distribution at the 0.1\% level. These results
are consistent with the view that large intrinsic variations of the CO
properties of the PNe (Huggins et al. 1996) dominate over evolutionary
effects for the ensemble as a whole. They also indicate that the presence
of a CO envelope is correlated with higher [\ion{N}{2}] emission in
the nebula. This is likely to be mainly a direct effect in which
[\ion{N}{2}] is enhanced in dense interface gas, or shocked gas often
associated with bipolar flows and dense tori. It may also be partly
indirect, in that higher mass progenitors, which often have an
enhanced N abundance, preferentially have more massive molecular
envelopes. These will be discussed in more detail in a separate paper.

\subsection{Radial Velocities}
  
The CO spectra of the 40 detected PNe provide a useful and homogeneous
set of PN radial velocities ($V_{\rm o}$ in Table 1). In the cases
where the CO spectrum is relatively clear of IS line contamination and
the S/N is good, the uncertainty in the measured velocities is small
$\sim 1$--2~\kms, and is limited by the intrinsic asymmetry in the
line profiles. Even in cases with severe IS contamination or low
S/N, the uncertainty is almost always $\la 10$~\kms.

High or medium quality (uncertainties $\la 10$~\kms) optical
measurements of the radial velocities of the detected PNe are
available in only 18 cases (see Section 4). For this set, the mean
value of $V_{\rm o} - V_{\rm opt} = 5$~\kms\ with an rms scatter of
12~\kms.  Thus the CO and optical measurements provide similar, robust
estimates of the systemic velocity, with part of the residual scatter
being real, reflecting differences in the structure of the ionized and
molecular gas in the PNe.  In the other 22 cases where CO is detected
but the optical uncertainties are large, $\ga 10$~\kms\ (10 PNe), or
are non-existent (12 PNe), the CO data provide the preferred or the
only available radial velocity measurements of the PNe.

\subsection{Line Profiles and Widths}

A striking feature of the CO spectra shown in Fig.~1 is the variety of
the line profiles. They can be broadly classified into 4 basic types:
(I) parabolic, similar to the CO spectra typically seen in the
precursor AGB envelopes; (II) flat-topped; (III) double-peaked; and
(IV) triangular, usually asymmetrical.  There are also intermediate
types: e.g., with a distinct but not deep concavity, between types II
and III; and with strongly asymmetrical double-peaks, between types
III and IV.

These different profiles probably arise from a combination of
effects. The first is the evolution of the structure of the envelope
from the relatively complete, spherically symmetric distribution seen
around AGB stars, to one in which the envelope has been hollowed out
by dissociating and ionizing radiation and/or shaped by the action of
jet outflows (see Section 1) into bi-cones, rings, tori, etc., which
may be oriented at different angles to the line of sight. Red-blue
asymmetries in the profiles then arise from back-front asymmetries in
the distribution of molecular gas.  A second effect is the likely
decrease in the opacity of the CO line as the CO is destroyed. A third
effect is the size of the PN cavity relative to the telescope beam
size. If the third effect were dominant, one would expect that the
profile shapes would be correlated with the angular diameter of the
PNe. 068.3$-$02.1 is among the smallest PNe (1.3\arcsec\ Aaquist \&
Kwok 1990), and is the case with the most parabolic profile,
consistent with the emission arising in a relatively undisturbed
envelope. However, there is no general correlation between the profile
shapes and the angular sizes of the ionized nebulae for the sample as a
whole. The prevalence of asymmetries, and strongly double-peaked
profiles even among small nebulae, suggest that the structural effects
in the envelopes are of major importance.

A second, striking feature of the CO spectra is the large widths of
the lines.  The distribution of $\Delta V$ listed in Table 1 is shown
in Fig.~2, where the lower limits for $\Delta V$ obtained from
Gaussian fits are included.  For comparison, Fig.~2 also shows the
distribution of CO line widths (twice the expansion velocity) of the
sample of 65 bright C-rich AGB stars reported by Olofsson et al.
(1993); the distribution of the widths of OH/IR stars is similar as
discussed in that paper.  The difference between PNe and AGB line
widths is clear from the figure. High velocity gas is rarely seen in
AGB envelopes (median width $= 25$~\kms), but is the norm in PNe
envelopes (median width $\ga 45$~\kms).

The expansion velocity of the ionized gas in PNe is also larger than
the expansion velocity of AGB envelopes, typically by a factor of
about two. The mean expansion velocities given for PNe in the catalog
of Weinberger (1989) are 20.3~\kms\ for [\ion{O}{3}] and 22.5~\kms\
for [\ion{N}{2}], and the corresponding line widths fall near the
center of the distribution of CO line widths shown in Fig.~2. In PNe
the large expansion velocities of the ionized gas are usually ascribed
to acceleration by the pressure in the nebulae caused by
photo-ionization or fast winds. In the case of the CO lines, the large
line widths of the PN spectra in Fig.~1 are mostly formed by extended
line wings, e.g., M 1-59 (023.9$-$02.3). Similar wings are commonly
seen in proto-PNe before the onset of ionization, and in the few cases
of young PNe where the CO line wings have been studied at high angular
resolution, e.g., M 1-16 (Huggins et al. 2000) and He 3-1475 (Huggins
et al. 2004), it is found that the wings arise from molecular gas
entrained in the sides of high velocity jets. Thus although the
kinematics of the molecular gas may be affected by thermal pressure in
the ionized gas, the frequency with which wings are seen in the CO
spectra reported here probably reflects the frequency of jets in PN
formation. 

In a few of the cases reported here, the whole CO line appears to be
broadened. The most extreme case is K~3-17 (039.8+02.1) which shows a
steep-sided, roughly flat-topped profile (with some IS contamination
near the center) that is 90~\kms\ wide. This profile is unique among
PNe and the object deserves further study.

\section{Conclusions}

This paper reports a sensitive survey of CO emission in PNe. 110 PNe
were observed and 40 were detected.  The success of the survey
demonstrates the usefulness of array receivers for observing CO in PNe
or other small sources lying close to the Galactic plane. The results
substantially extend the available data on the molecular gas in PNe
for statistical study, and provide individual cases for further
detailed observation.

\acknowledgements
We thank the staff of the IRAM 30~m telescope for help with the
observations.  This work has been supported in part by NSF grants AST
99-86159 and AST 03-07277 (PJH), and by the Spanish MEC under grants
AYA2003-7584 (RB), ESP2003-04957 and FEDER grants (PP).


\clearpage


\clearpage





\begin{figure}
\figurenum{1}
\includegraphics[scale=.7,angle=-90]{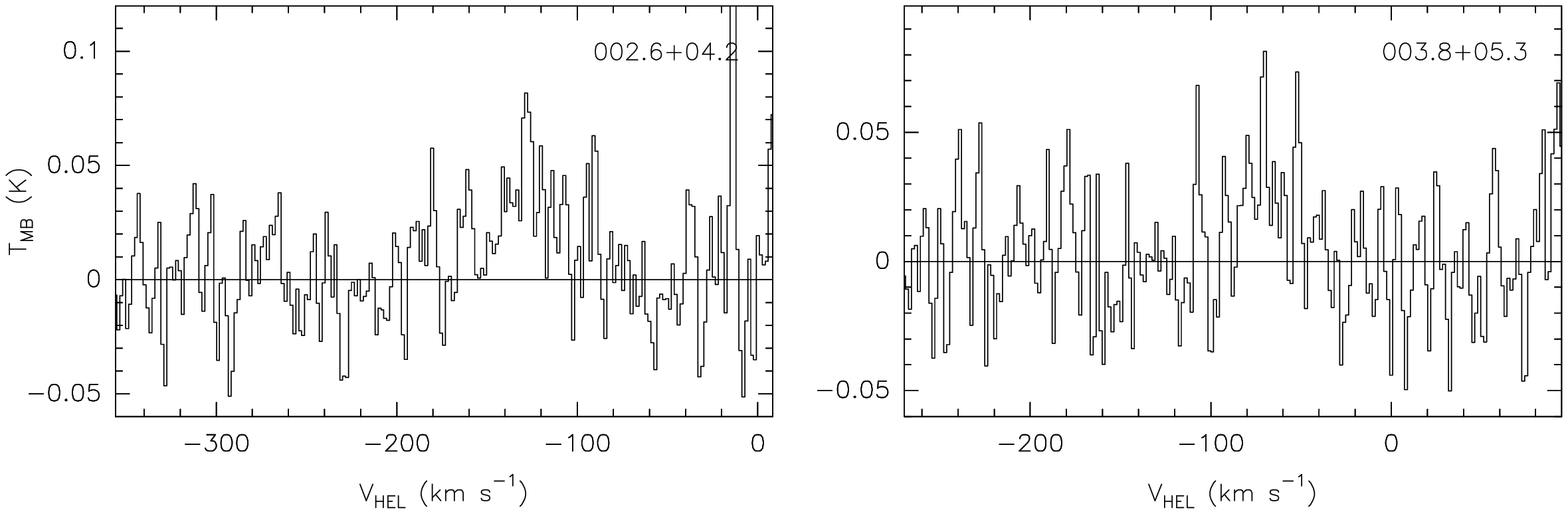}

\vspace{0.6cm}
\includegraphics[scale=.7,angle=-90]{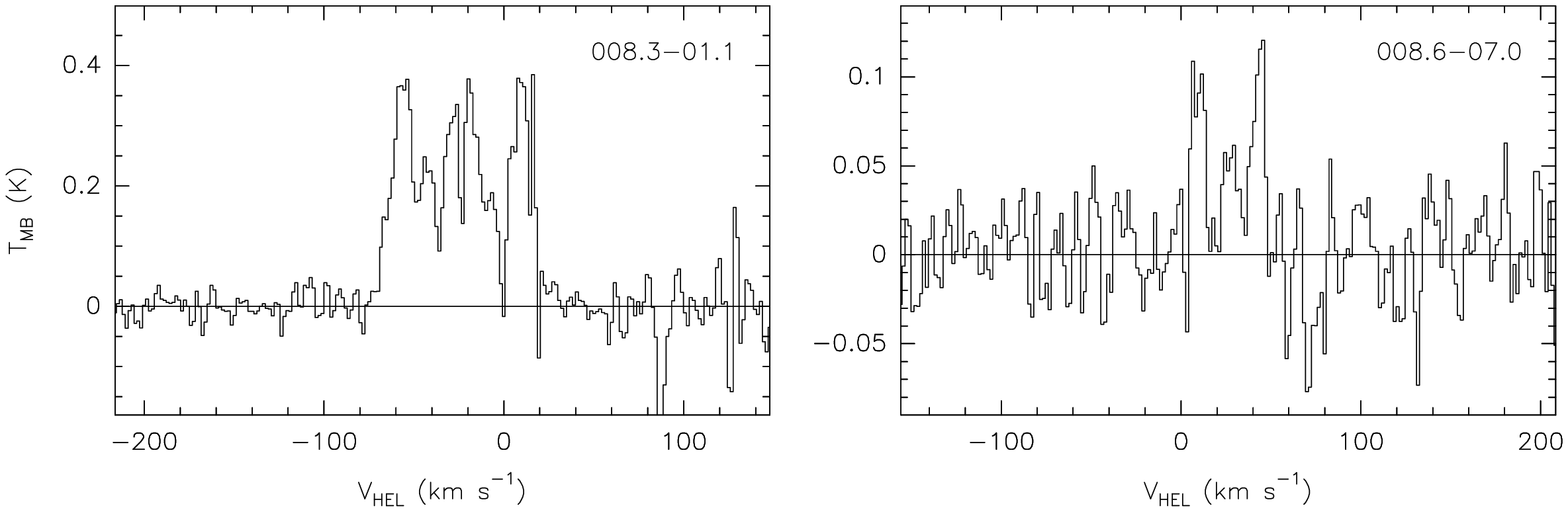}
\caption[]{CO $J=2-1$ spectra of the PNe detected in the
survey. The PNe are identified by the galactic name in each panel and
are ordered by Galactic longitude. See section 4 for comments on
individual spectra.  }
\end{figure}
\clearpage
\begin{figure}
\figurenum{1}
\includegraphics[scale=.7,angle=-90]{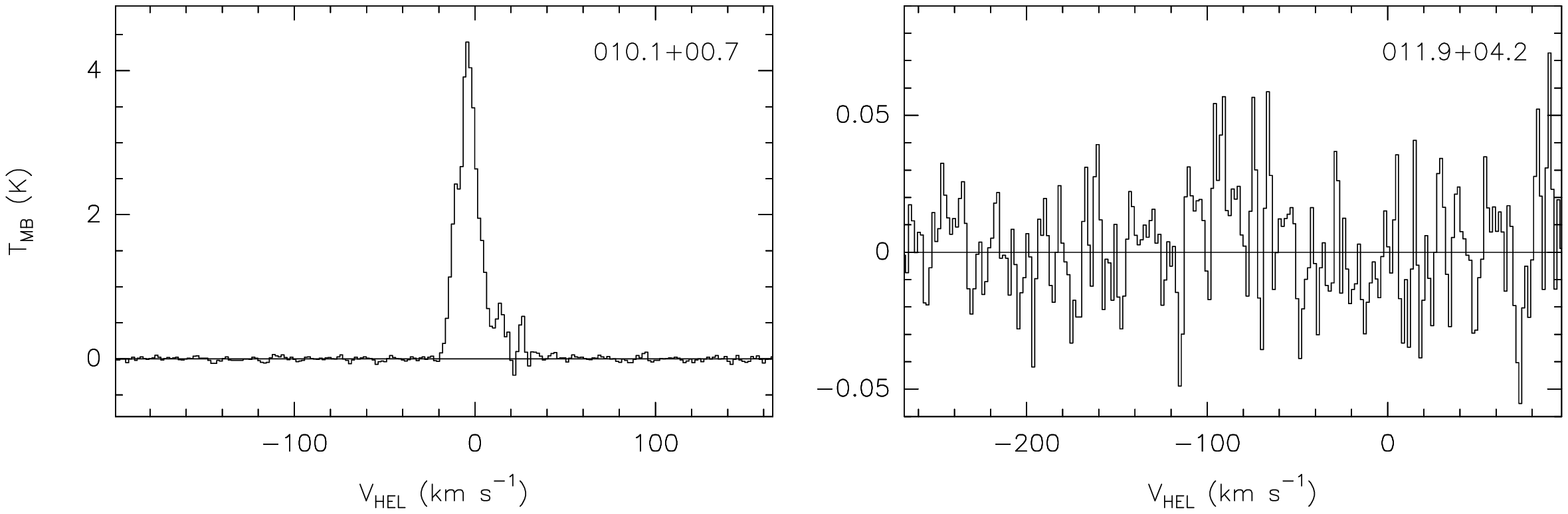}

\vspace{0.6cm}
\includegraphics[scale=.7,angle=-90]{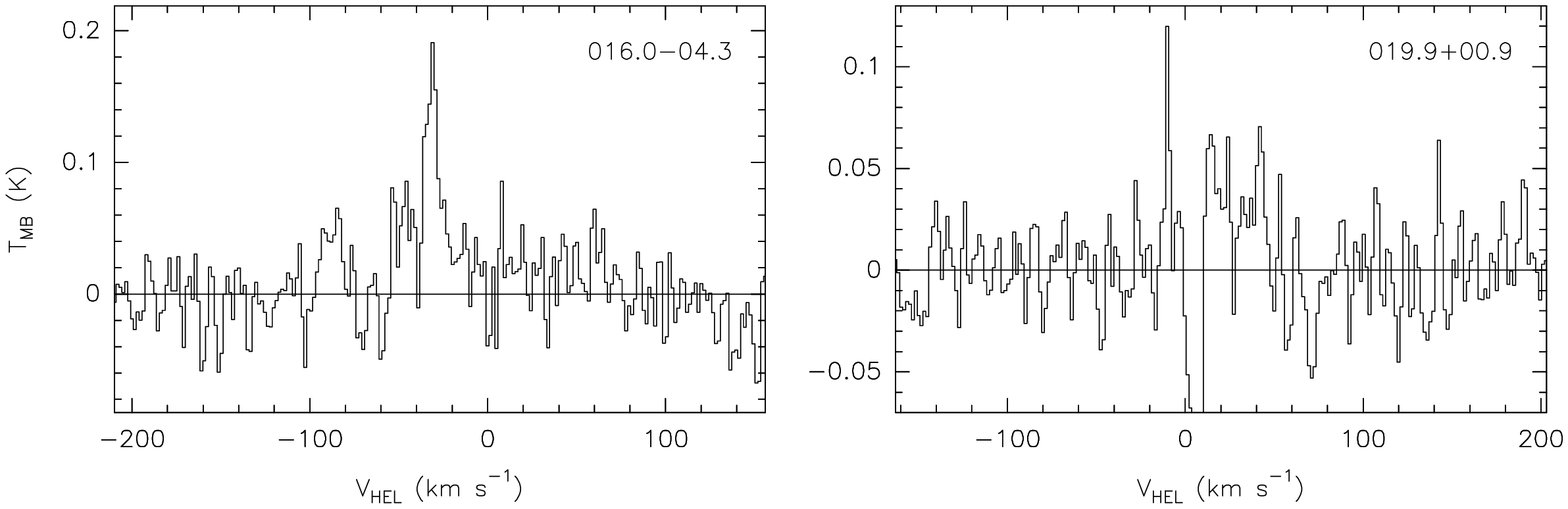}
\caption{continued}
\end{figure}
\clearpage
\begin{figure}
\figurenum{1}
\includegraphics[scale=.7,angle=-90]{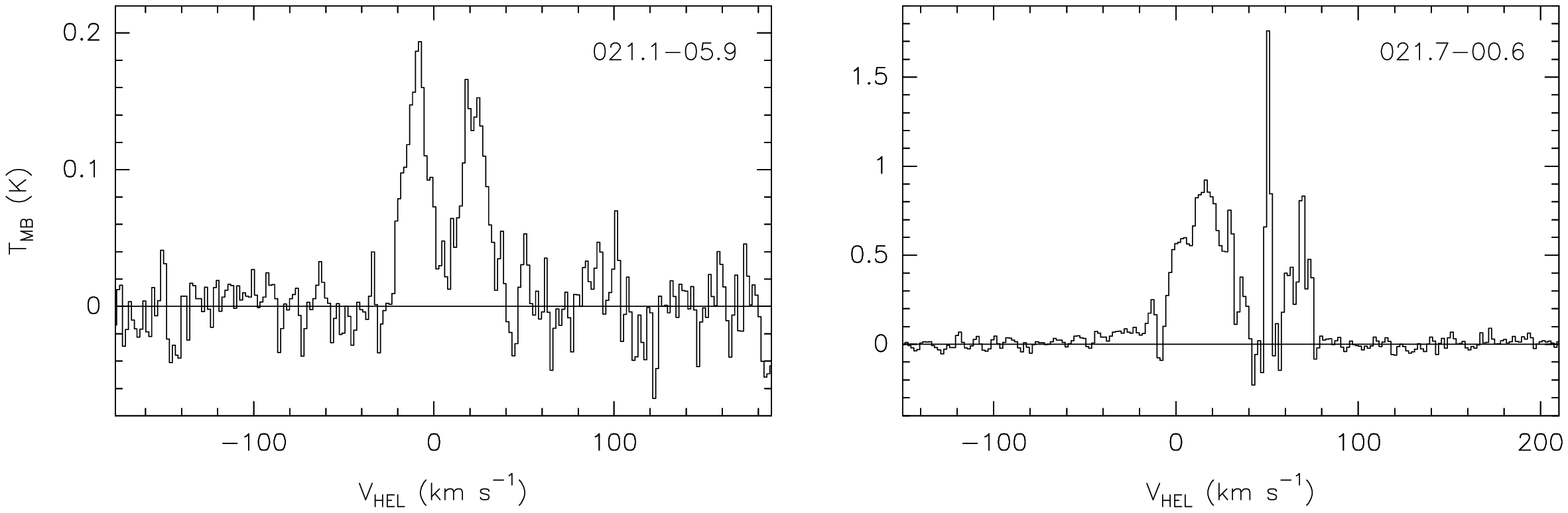}

\vspace{0.6cm}
\includegraphics[scale=.7,angle=-90]{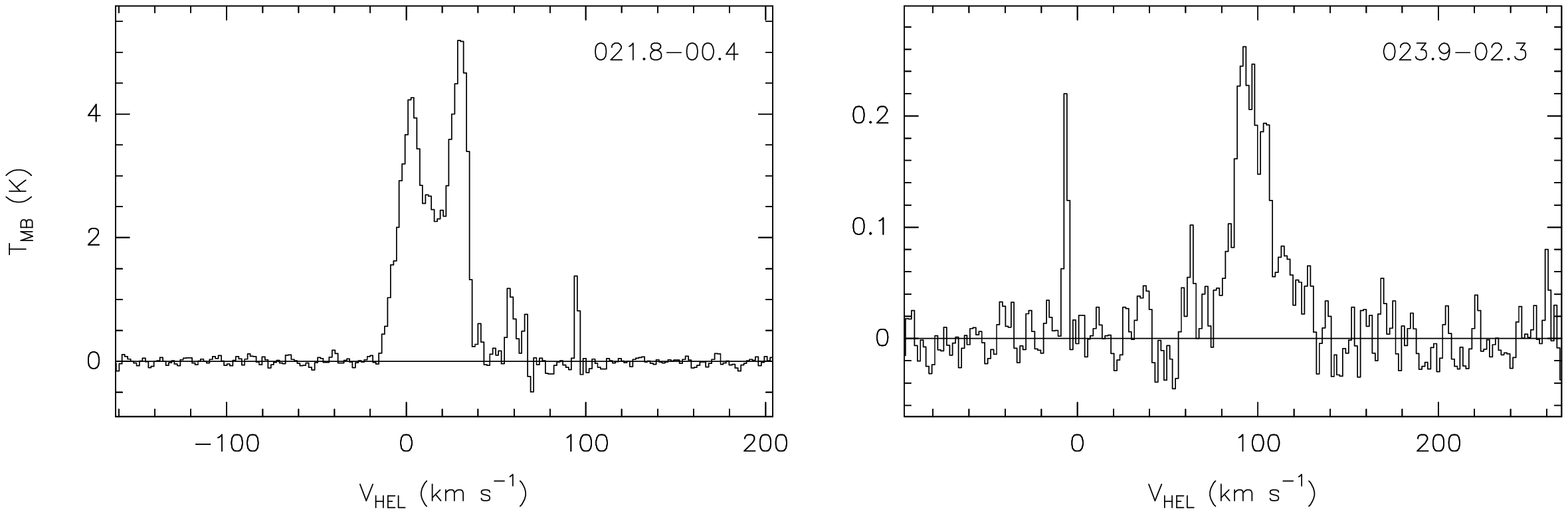}
\caption{continued}
\end{figure}
\clearpage
\begin{figure}
\figurenum{1}
\includegraphics[scale=.7,angle=-90]{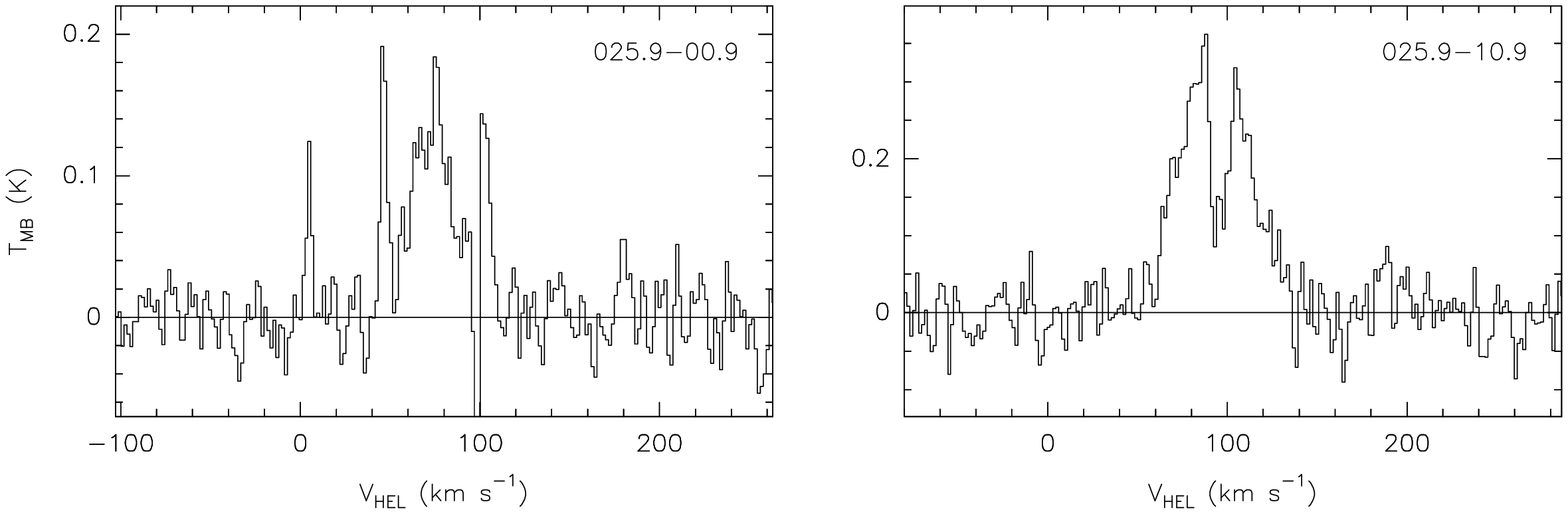}

\vspace{0.6cm}
\includegraphics[scale=.7,angle=-90]{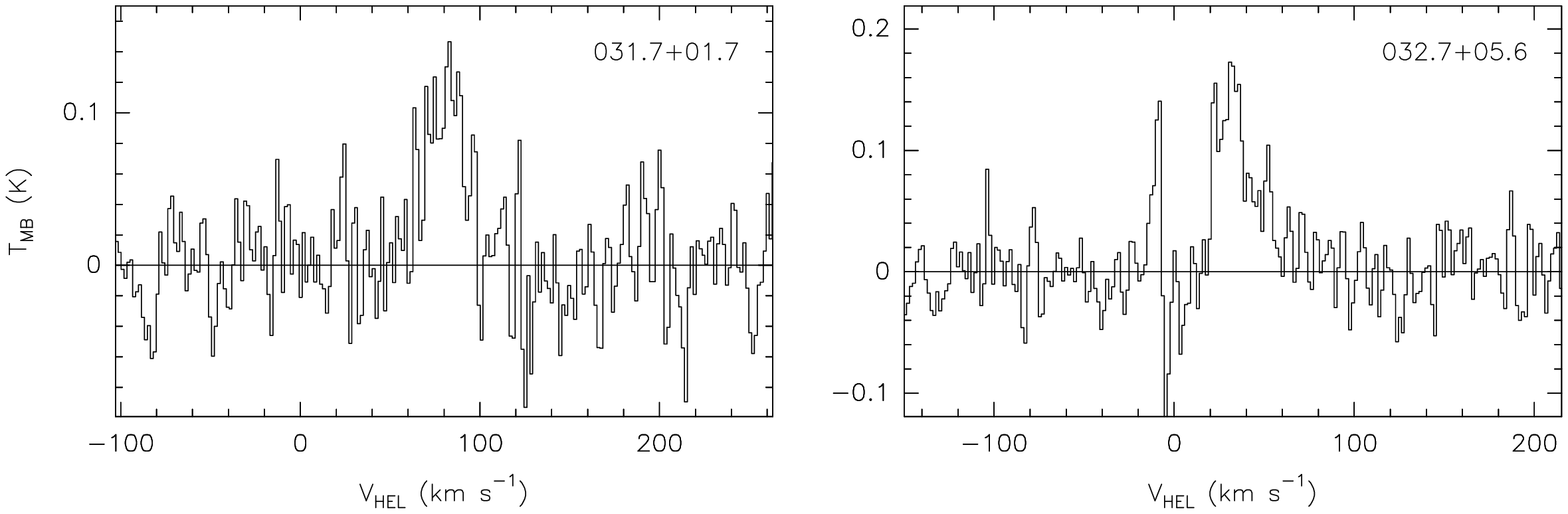}
\caption{continued}
\end{figure}
\clearpage
\begin{figure}
\figurenum{1}
\includegraphics[scale=.7,angle=-90]{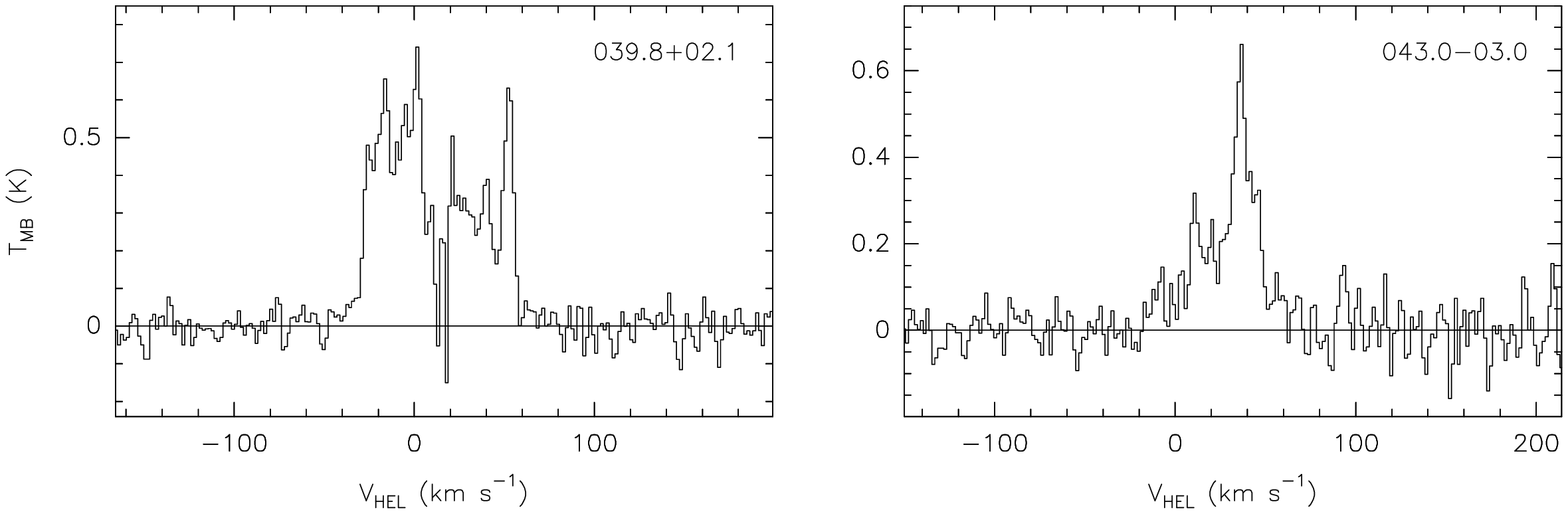}

\vspace{0.6cm}
\includegraphics[scale=.7,angle=-90]{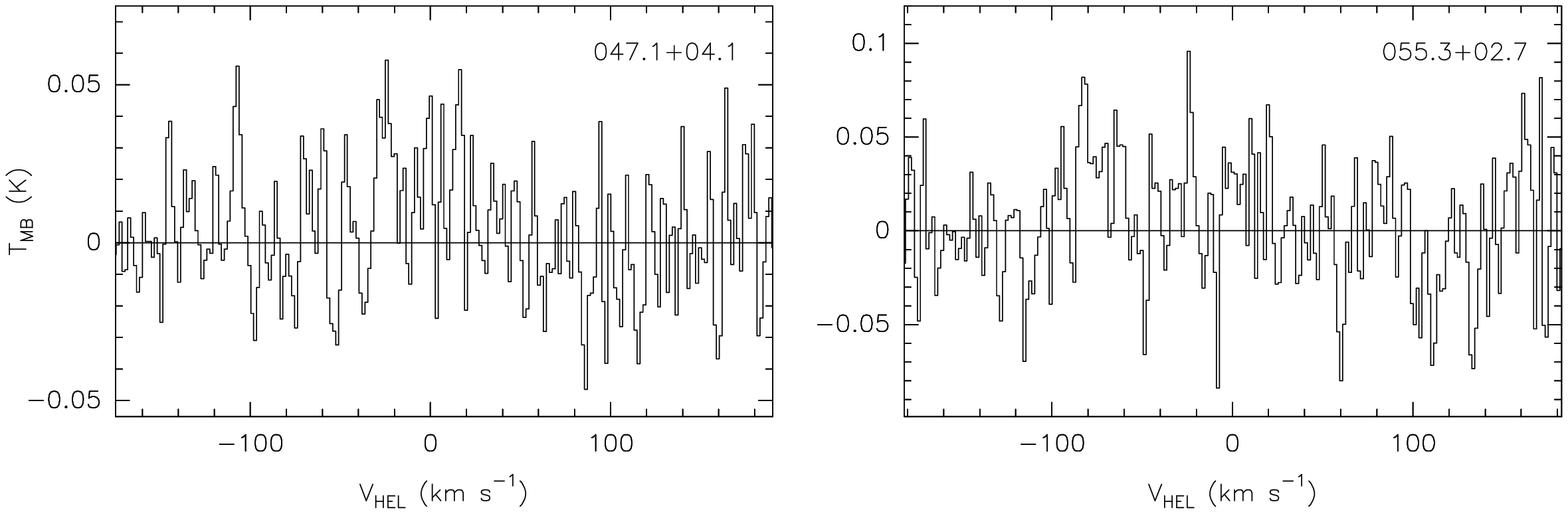}
\caption{continued}
\end{figure}
\clearpage
\begin{figure}
\figurenum{1}
\includegraphics[scale=.7,angle=-90]{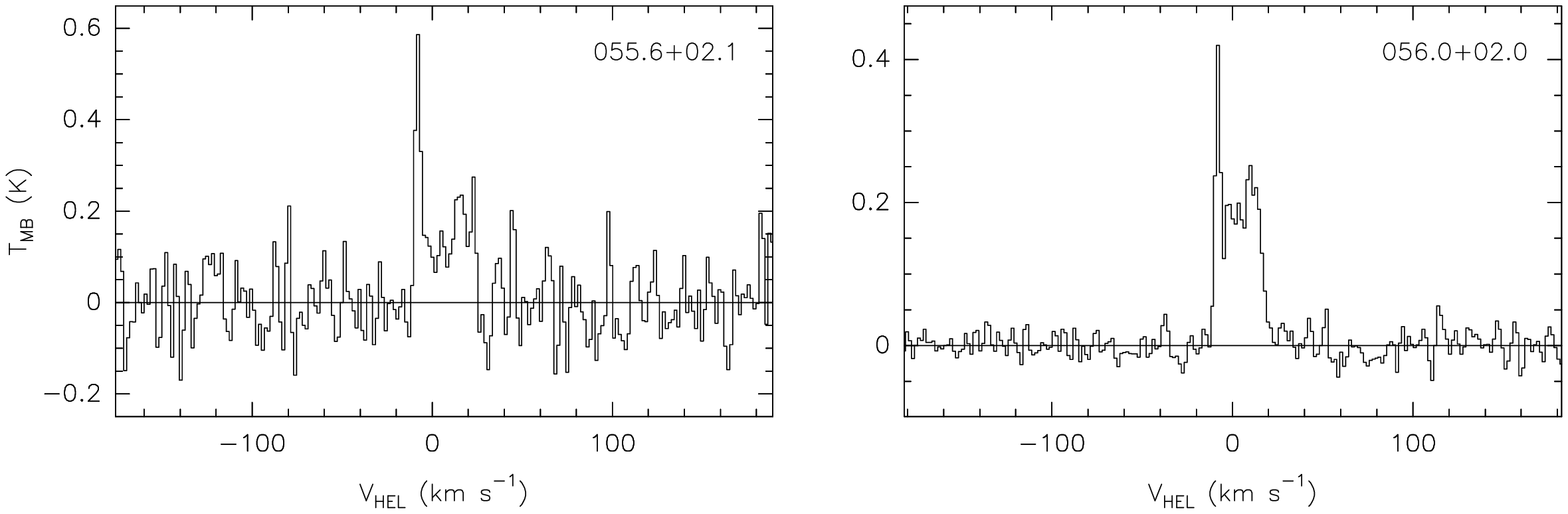}

\vspace{0.6cm}
\includegraphics[scale=.7,angle=-90]{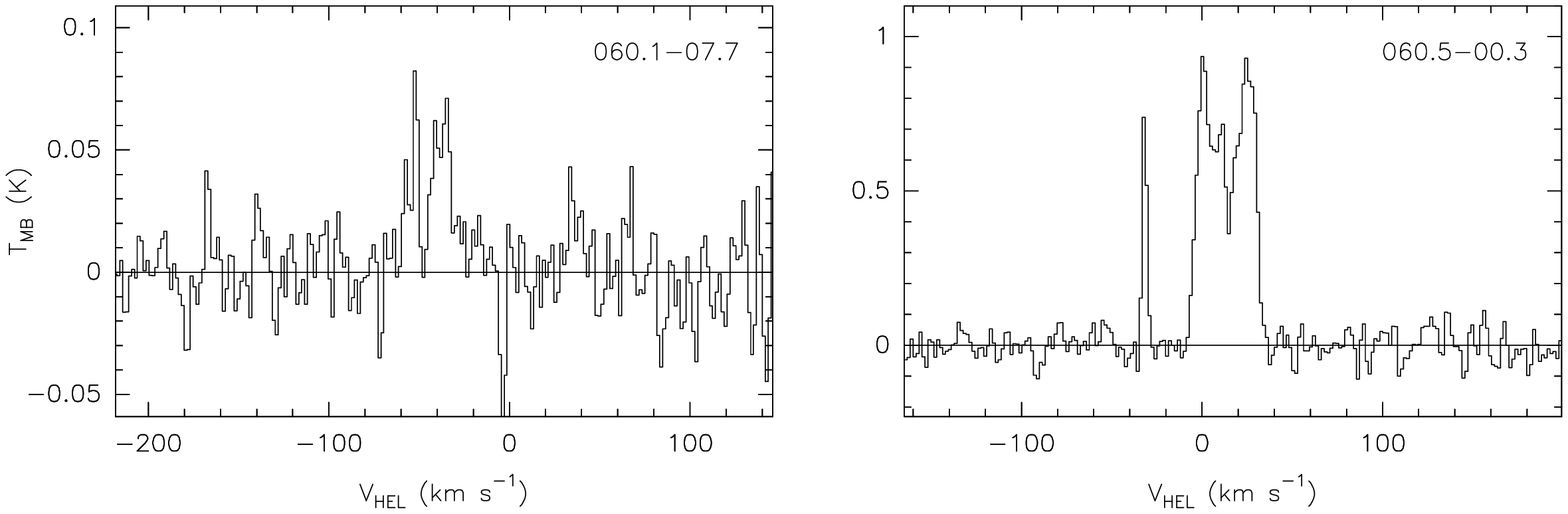}
\caption{continued}
\end{figure}
\clearpage
\begin{figure}
\figurenum{1}
\includegraphics[scale=.7,angle=-90]{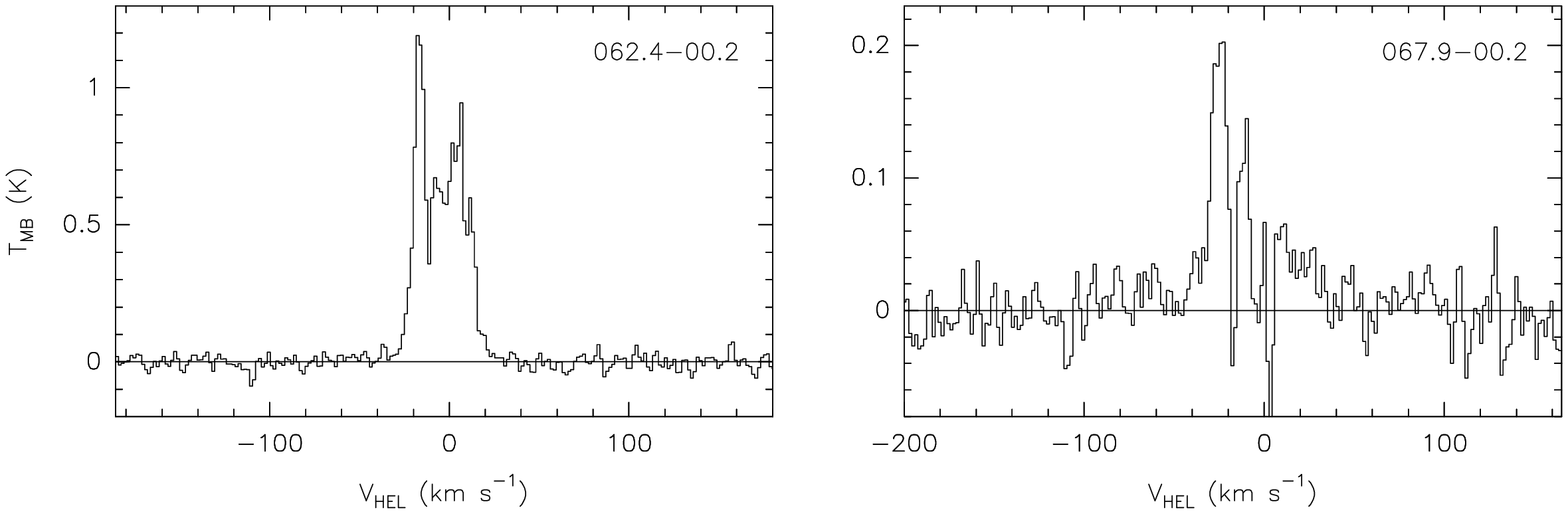}

\vspace{0.6cm}
\includegraphics[scale=.7,angle=-90]{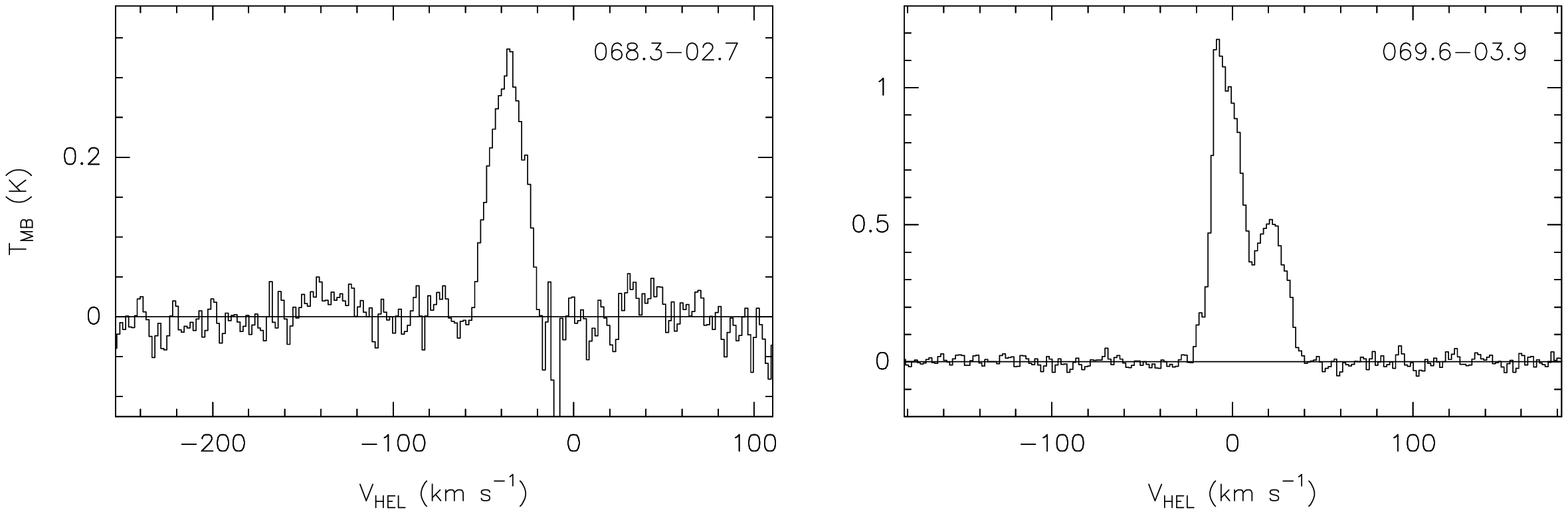}
\caption{continued}
\end{figure}
\clearpage
\begin{figure}
\figurenum{1}
\includegraphics[scale=.7,angle=-90]{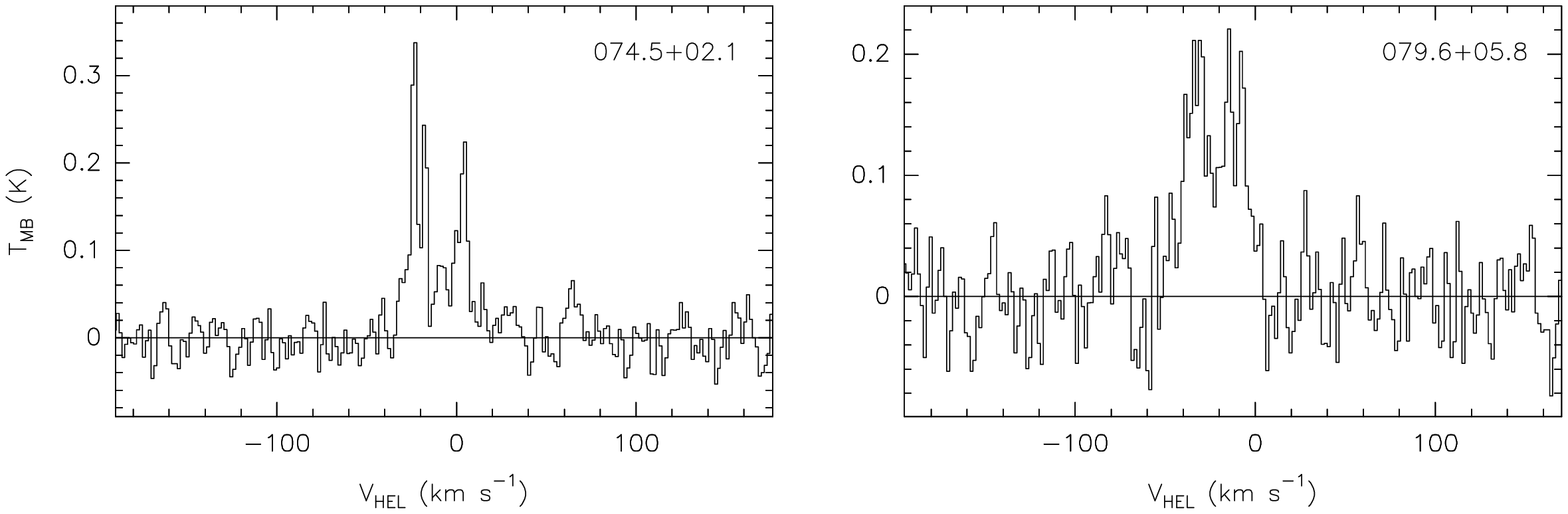}

\vspace{0.6cm}
\includegraphics[scale=.7,angle=-90]{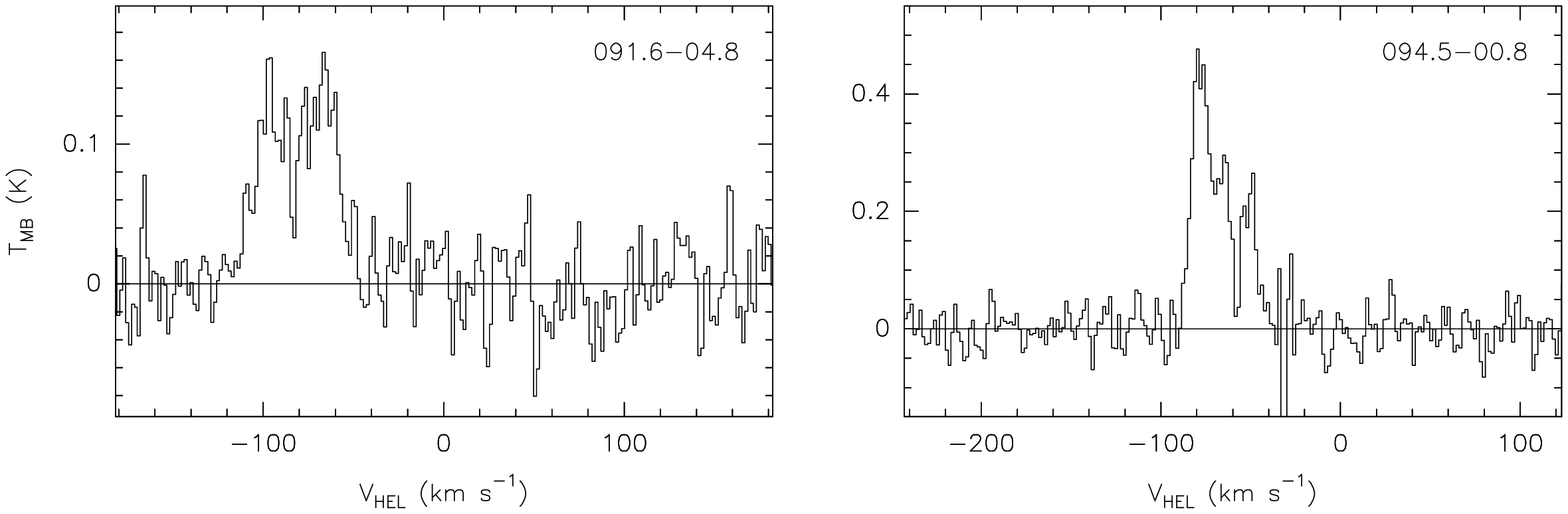}
\caption{continued}
\end{figure}
\clearpage
\begin{figure}
\figurenum{1}
\includegraphics[scale=.7,angle=-90]{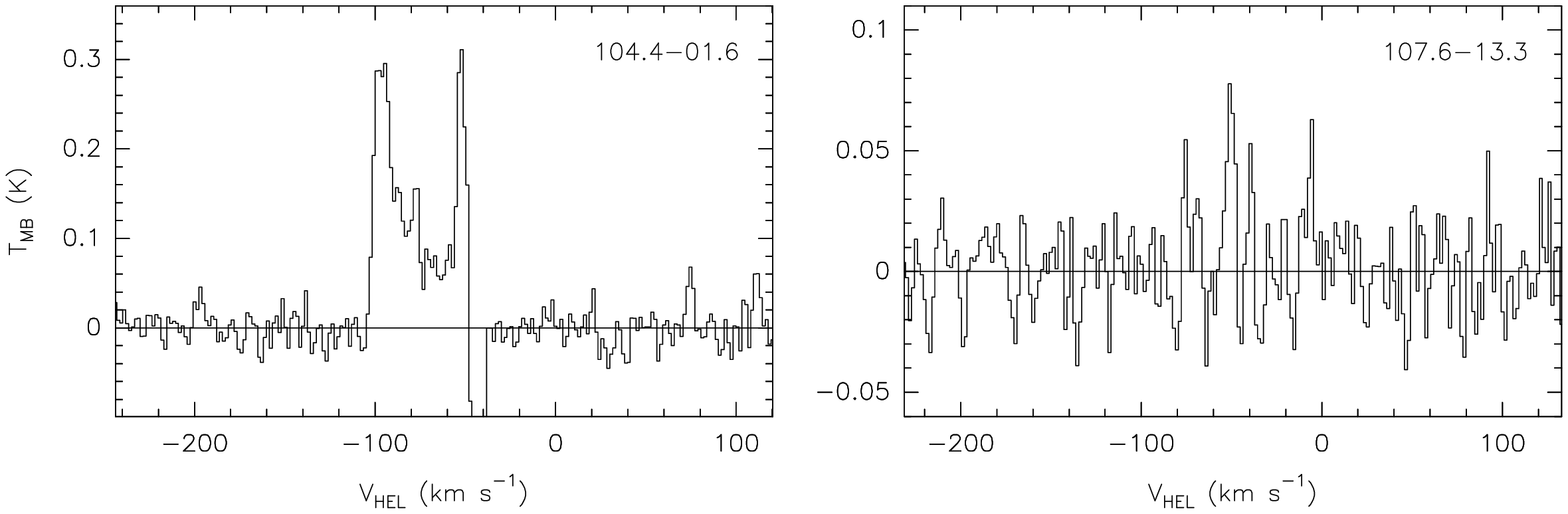}

\vspace{0.6cm}
\includegraphics[scale=.7,angle=-90]{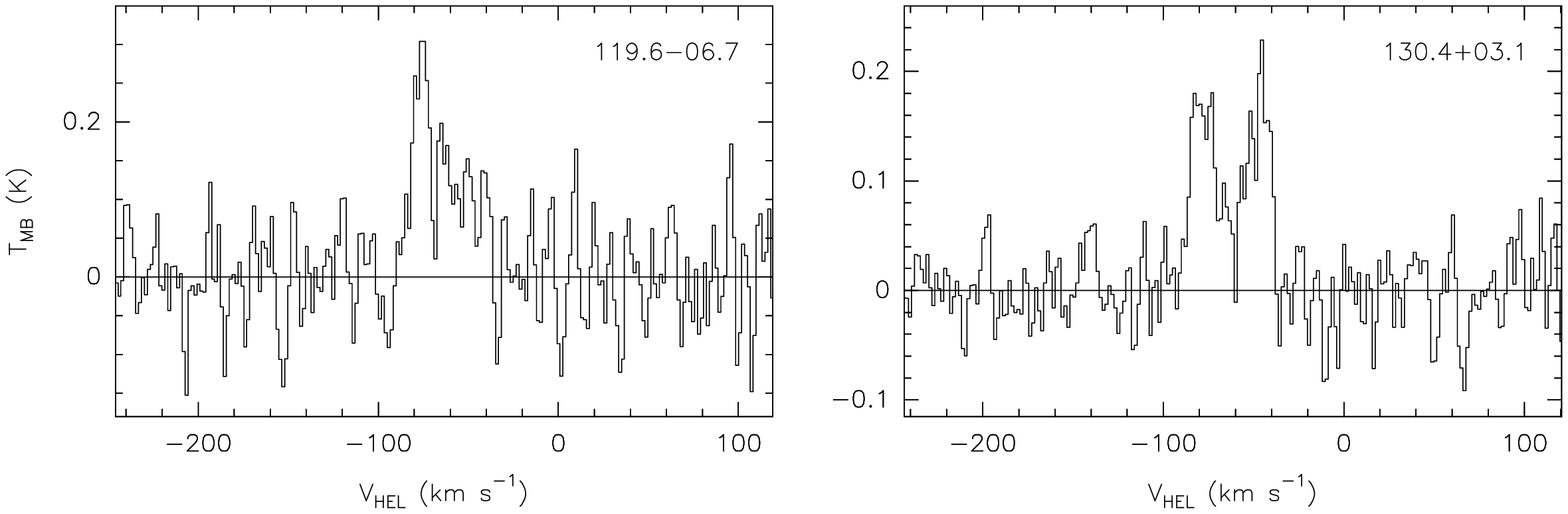}
\caption{continued}
\end{figure}
\clearpage
\begin{figure}
\figurenum{1}
\includegraphics[scale=.7,angle=-90]{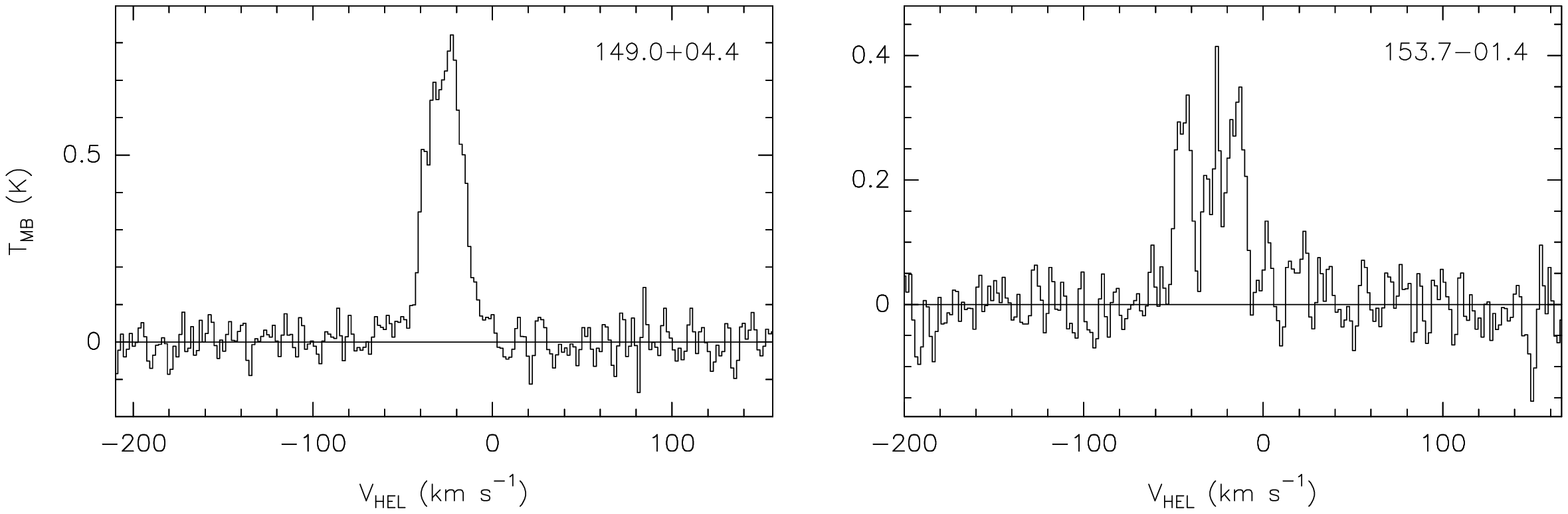}

\vspace{0.6cm}
\includegraphics[scale=.7,angle=-90]{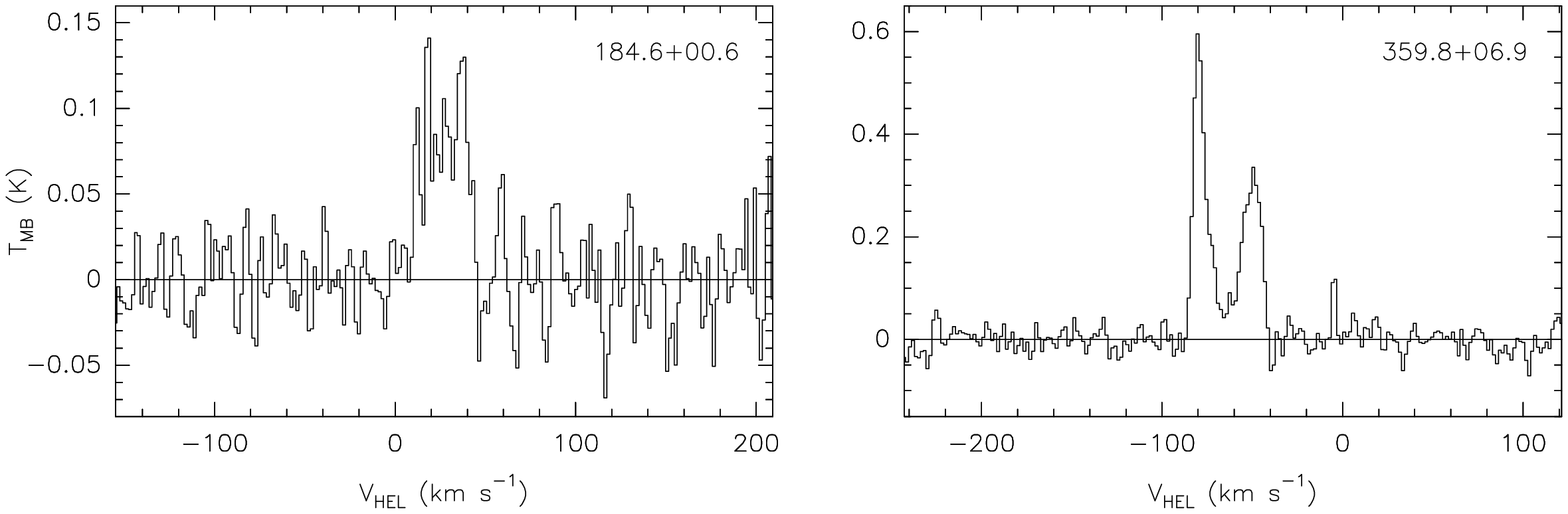}
\caption{continued}
\end{figure}
\clearpage

\begin{figure}
\figurenum{2}
\plotone{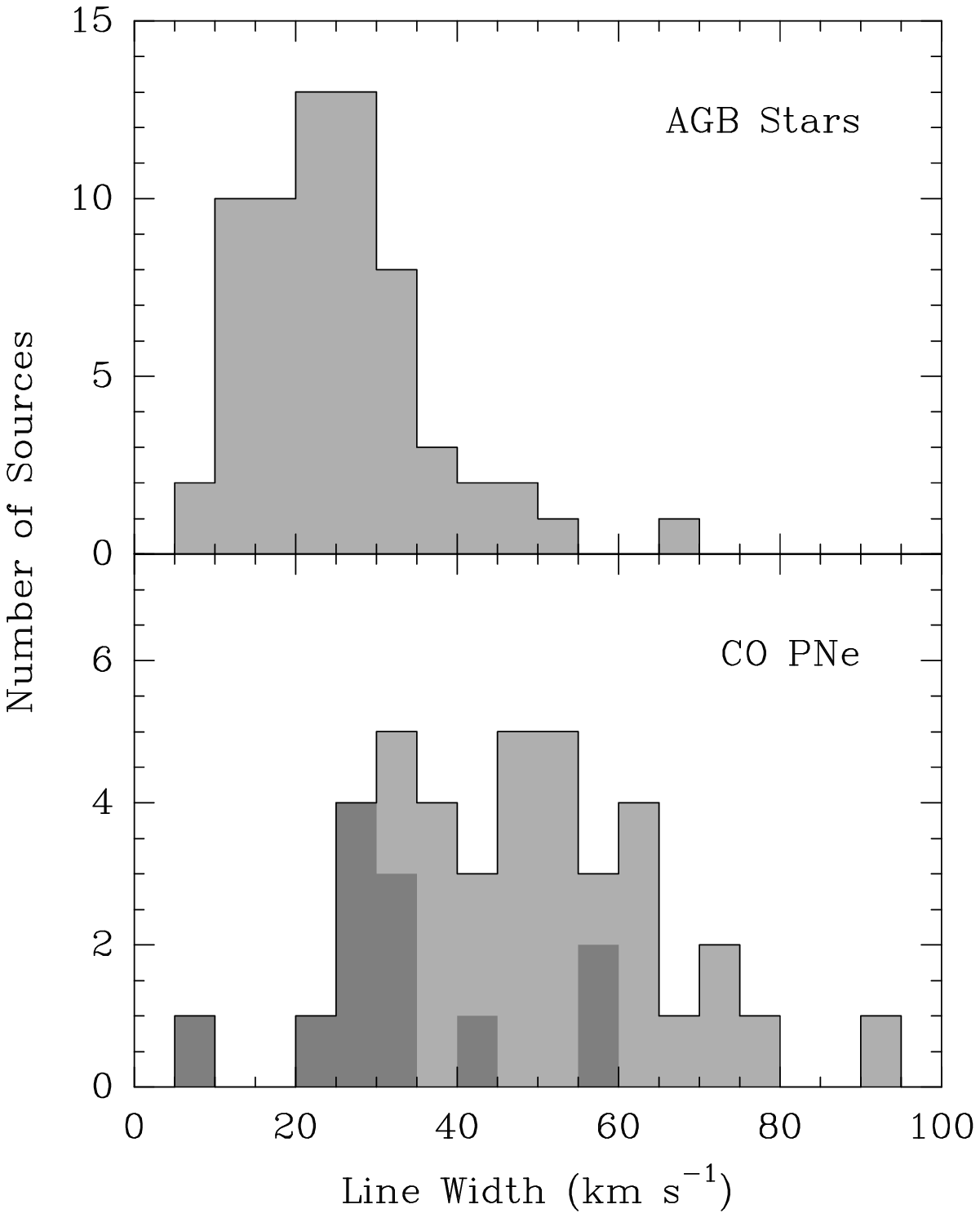}
\caption[]{Line width comparison of PNe and AGB stars. \emph{Lower
panel}: Distribution of the CO line widths of
the PNe detected in the survey. The darker regions denote lower limits
to the CO widths. \emph{Upper panel}: Distribution of the CO line
widths of 65 bright carbon stars. Data from Olofsson et
al.\ (1993).
}
\end{figure}
\clearpage

\begin{deluxetable}{llrrrrrrl}
\tabletypesize{\footnotesize} 
\tablecolumns{9}
\tablewidth{0pc}
\tablecaption{CO Survey of Young Planetary Nebulae}
\tablehead{
\colhead{PN} & \colhead{Name} & \colhead{rms} & \colhead{ $I$ } & \colhead{$V_{\rm o}$}   & \colhead{$\Delta\, V$} & \colhead{$T_{\rm p}$} & \colhead{$V_{\rm IS}$} & \colhead{Comments} \\
\colhead{} & \colhead{}       & \colhead{ (mK) } & \colhead{ (K\,km\,s$^{-1}$) } & \colhead{(km\,s$^{-1}$) }  & \colhead{(km\,s$^{-1}$)} & \colhead{(mK)}\ & \colhead{(km\,s$^{-1}$)} & } 
\startdata
 000.1+17.2 & PC 12    & 25  & \nodata & \nodata & \nodata & \nodata
&\nodata     &  n, g1  \\ 
 000.7+04.7 & H 2-11   & 47  & \nodata & \nodata & \nodata & \nodata &
$-$12 +11  &  n, g1 \\ 
 002.6+02.1 & Te 1580  & 43  & \nodata & \nodata & \nodata & \nodata &
$-$18 $-$4 &  n, g3 \\  
 002.6+04.2 & Th 3-27  & 21  & 2.4     &  $-$127 &   $>$55 &      40
&$-$17 +9    &  t, g2 \\ 
 002.8+01.7 & H 2-20   & 27  & \nodata & \nodata & \nodata & \nodata &
$-$20 $-$3 &  n, g1 \\ 
 002.8+01.8 & Te 1567  & 30  & \nodata & \nodata & \nodata & \nodata &
$-$15 $-$4, +15 +19 & n, g2 \\ 
 003.6+03.1 & M 2-14   & 29  & \nodata & \nodata & \nodata & \nodata &
$-$6 $-$3   &  n, g1 \\ 
 003.8+05.3 & H 2-15   & 24  & 1.2     &   $-$69 &   $>$32 & 36      &
\nodata     &  t, g2  \\ 
 004.9+04.9 & M 1-25   & 45  & \nodata & \nodata & \nodata & \nodata &
+44 +48, +94 +100 & n, g1  \\  
 006.0+03.1 & M 1-28   & 33  & \nodata & \nodata & \nodata & \nodata &
$-$22 $-$12  &  n, g4 \\ 
 006.3+03.3 & H 2-22   & 32  & \nodata & \nodata & \nodata & \nodata &
$-$57 $-$36, $-$16 $-$5 & n, g1 \\ 
 006.4+02.0 & M 1-31   & 41  & \nodata & \nodata & \nodata & \nodata &
$-$5 +1    &  n, g3 \\ 
 006.5$-$03.1 & H 1-61   & 35  & \nodata & \nodata & \nodata & \nodata
& $-$3 +2    & n, g1  \\ 
 006.7$-$02.2 & M 1-41   & 24  & \nodata & \nodata & \nodata & \nodata
& $-$100 +25  &  n, g2, s  \\ 
 007.2+01.8 & Hb 6     & 30  & \nodata & \nodata & \nodata & \nodata &
+3 +12    & n, g3  \\ 
 007.8$-$04.4 & H 1-65   & 27  & \nodata & \nodata & \nodata & \nodata
& \nodata   &  n, g1  \\ 
 008.3$-$01.1 & M 1-40   & 22  & 15.8    &   $-$35 &      70 & 360
& $-$25 +20, +84 +145  &  d, g2, s  \\ 
 008.3$-$07.3 & NGC 6644 & 27  & \nodata & \nodata & \nodata & \nodata
& \nodata   &  n, g3  \\ 
 008.6$-$07.0 & He 2-406 & 26  & 2.4     &     +27 &      50 &  100
& \nodata   & d, g2  \\ 
 010.1+00.7 & NGC 6537 & 31  & 59.7    &    $-$4 &      30 &  4340   &
+1 +34, +88 +95 &  d, g2 \\ 
 010.7$-$06.4 & IC 4732  & 24  & \nodata & \nodata & \nodata & \nodata
& \nodata & n, g3  \\ 
 011.0+05.8 & NGC 6439 & 46  & \nodata & \nodata & \nodata & \nodata &
\nodata & n, g3  \\ 
 011.1+11.5 & M 2-13   & 49  & \nodata & \nodata & \nodata & \nodata &
\nodata & n, g3  \\ 
 011.9+04.2 & M 1-32   & 19  & 0.77    &   $-$90 &   $>$27 &    27   &
$-$10 $-$6 &  t, g1 \\ 
 012.6$-$02.7 & M 1-45   & 31  & \nodata & \nodata & \nodata & \nodata
& +41 +63 &  n, g1  \\ 
 013.4$-$03.9 & M 1-48   & 41  & \nodata & \nodata & \nodata & \nodata
& \nodata & n, g2 \\ 
 014.9$-$03.1 & SaSt 3-16& 29  & \nodata & \nodata & \nodata & \nodata
& \nodata & n, g1  \\ 
 015.9+03.3 & M 1-39   & 23  & \nodata & \nodata & \nodata & \nodata &
+0 +9   &  n, g1  \\ 
 016.0$-$04.3 & M 1-54   & 26  & 2.8     &   $-$32 &   $>$23 & 110
& \nodata & d, g2  \\ 
 016.4$-$01.9 & M 1-46   & 42  & \nodata & \nodata & \nodata & \nodata
& +23 +45 & n, g1 \\ 
 018.9+04.1 & M 3-52   & 32  & \nodata & \nodata & \nodata & \nodata &
$-$6 $-$1 & n, g2  \\ 
 019.7$-$04.5 & M 1-60   & 28  & \nodata & \nodata & \nodata & \nodata
& $-$14 +0 & n, g3   \\ 
 019.9+00.9 & M 3-53   & 19  & 1.8     &     +18 &   $>$41 &      40 &
$-$10 +10, +22 +26 & d, g2  \\ 
 020.9$-$01.1 & M 1-51   & 43  & \nodata & \nodata & \nodata & \nodata
& $-$14 +80 & n, g4, s \\ 
 021.1$-$05.9 & M 1-63   & 22  & 5.6     &      +6 &      60 & 170
& \nodata & d, g2  \\ 
 021.7$-$00.6 & M 3-55   & 31  & 30.0    &     +16 &      46 &   838
& $-$17 $-$4, +23 +71 &  d, g1, s  \\ 
 021.8$-$00.4 & M 3-28   & 62  &  147.1  &     +17 &      52 & 5160
& $-$11 +99  & d, g2  \\ 
 022.1$-$02.4 & M 1-57   & 33  & \nodata & \nodata & \nodata & \nodata
& $-$9 $-$3, +17 +21  &  n, g2  \\ 
 023.3$-$07.6 & MaC 1-16 & 43  & \nodata & \nodata & \nodata & \nodata
& \nodata  & n, g2 \\ 
 023.9$-$02.3 & M 1-59   & 21  & 6.0     &     +97 &      63 &    220
& $-$8 $-$2 &  d, g3  \\ 
 024.1+03.8 & M 2-40   & 22  & \nodata & \nodata & \nodata & \nodata &
$-$13 $-$2  &  n, g1  \\ 
 024.3$-$03.3 & Pe 1-17  & 28  & \nodata & \nodata & \nodata & \nodata
& $-$8 $-$1, +38 +45  & n, g2 \\ 
 024.8$-$02.7 & M 2-46   & 29  & \nodata & \nodata & \nodata & \nodata
& $-$11 $-$2, +40 +49 & n, g1  \\ 
 025.9$-$00.9 & Pe 1-14  & 20  & 4.2     &     +74 &   $>$32 & 124
& +5 +49, +74 +105 & d, g2, s \\ 
 025.9$-$10.9 & Na 2     & 33  & 13.5    &     +95 &      73 & 290     &
\nodata &  d, g2  \\ 
 028.5+05.1 & K 3-2   & 29  & \nodata & \nodata & \nodata & \nodata &
$-$17 $-$14 & n, g1  \\ 
 031.7+01.7 & PC 20    & 31  & 3.2     &     +81 &   $>$26 &   116   &
$-$15 $-$1     & d, g3 \\ 
 032.5$-$03.2 & K 3-20   & 40  & \nodata & \nodata & \nodata & \nodata
& $-$10 $-$5  &   n, g1   \\ 
 032.7+05.6 & K 3- 4   & 26  & 3.5     &     +33 &      36 &  144    &
$-$18 +0 & d, g3  \\ 
 034.0+02.2 & K 3-13   & 42  & \nodata & \nodata & \nodata & \nodata &
$-$5 +1 & n, g3  \\ 
 036.9$-$01.1 & HaTr 11  & 41  & \nodata & \nodata & \nodata & \nodata
& $-$6 +3 & n, g4  \\ 
 038.2+12.0 & Cn 3-1   & 29  & \nodata & \nodata & \nodata & \nodata &
\nodata  &  n, g1  \\ 
 038.4$-$03.3 & K 4-19   & 40  & \nodata & \nodata & \nodata & \nodata
& $-$11 $-$5 &  n, g1 \\ 
 039.8+02.1 & K 3-17   & 39  & 34.2    &     +15 &      90 & 610     &
$-$1 +22 & d, g3 \\ 
 042.0+05.4 & K 3-14   & 36  & \nodata & \nodata & \nodata & \nodata &
\nodata  & n, g4 \\ 
 043.0$-$03.0 & M4-14    & 52  & 14.5    &     +24 &      76 &  610
& $-$77 $-$7 & d, g2  \\ 
 043.1+03.8 & M 1-65   & 60  & \nodata & \nodata & \nodata & \nodata &
$-$1 +16   & n, g1 \\ 
 045.9$-$01.9 & K 3-33   & 35  & \nodata & \nodata & \nodata & \nodata
& $-$14 $-$7, +7 +14 &  n, g1  \\ 
 047.1+04.1 & K 3-21   & 17  & 1.3     &   $-$4  &   $>$58 &   21    &
\nodata &  t, g2  \\ 
 048.1+01.1 & K 3-29   & 73  & \nodata & \nodata & \nodata & \nodata &
+13 +21   &  n, g3  \\ 
 048.7+01.9 & He 2-429 & 21  & \nodata & \nodata & \nodata & \nodata &
$-$73 $-$60, $-$14 $-$1 & n, g3  \\ 
 049.4+02.4 & He 2-428 & 38  & \nodata & \nodata & \nodata & \nodata &
$-$14 $-$3 &  n, g1  \\ 
 051.0+03.0 & He 2-430 & 40  & \nodata & \nodata & \nodata & \nodata &
$-$6 +3 & n, g3  \\ 
 051.9$-$03.8 & M 1-73   & 24  & \nodata & \nodata & \nodata & \nodata
& $-$1 +2  &  n, g1 \\ 
 052.5$-$02.9 & Me 1-1   & 25  & \nodata & \nodata & \nodata & \nodata
& \nodata &     n, g3  \\ 
 055.3+02.7 & He 1-1 & 32  &    1.3  &   $-$77 &   $>$26 &      47 &
$-$73 $-$70, $-$11 $-$8 &  t, g2 \\ 
 055.5$-$00.5 & M 1-71   & 51  & \nodata & \nodata & \nodata & \nodata
& +17 +27 &  n, g3  \\ 
 055.6+02.1 & He 1-2 & 74  &    5.0  &     +7  &      35 &  200    &
$-$12 $-$6 & d, g1  \\ 
 056.0+02.0 & K 3-35   & 18  &   5.7   &     +3  &      31 &  220    &
$-$13 $-$6 &  d, g2 \\ 
 057.9$-$01.5 & He 2-447 & 28  & \nodata & \nodata & \nodata & \nodata
& $-$3 +2, +17 +20 & n, g1  \\ 
 058.3$-$10.9 & IC 4997  & 19  & \nodata & \nodata & \nodata & \nodata &
\nodata  &   n, g1  \\ 
 059.0+04.6 & K 3-34   & 32  & \nodata & \nodata & \nodata & \nodata &
\nodata  &  n, g2  \\ 
 059.9+02.0 & K 3-39   & 92  & \nodata & \nodata & \nodata & \nodata &
$-$13 $-$8 & n, g1  \\ 
 060.1$-$07.7 & NGC 6886 & 16  &     1.4 &   $-$42 &   $>$28 &    46
& $-$6 $-$2 &  d, g3 \\ 
 060.5$-$00.3 & K 3-45   & 46  &    25.6 &     +15 &      43 &     870
& $-$35 $-$28, +14 +27 &  d, g1 \\ 
 060.5+01.8 & He 2-440 & 33  & \nodata & \nodata & \nodata & \nodata &
\nodata  &  n, g1  \\ 
 061.3+03.6 & He 2-437 & 24  & \nodata & \nodata & \nodata & \nodata &
\nodata  & n, g1  \\ 
 062.4$-$00.2 & M 2-48   & 26  &    25.1 &    $-$4 &      51 &    1170
& $-$13 +13  & d, g2  \\ 
 066.9+02.2 & K 4-37   & 26  & \nodata & \nodata & \nodata & \nodata &
$-$1 +5  &  n, g4  \\ 
 067.9$-$00.2 & K 3-52   & 19  &      4.7&   $-$25 &      67 &     190
& $-$29 +6     &    d, g1  \\ 
 068.3$-$02.7 & He 2-459 & 24  &     7.2 &   $-$37 &      35 &     330
& $-$17 $-$2  &  d, g1 \\ 
 068.8$-$00.0 & M 1-75   & 26  & \nodata & \nodata & \nodata & \nodata
& $-$29 +2  & n, g4, s  \\ 
 069.2+02.8 & K 3-49   & 37  & \nodata & \nodata & \nodata & \nodata &
\nodata  &   n, g1  \\ 
 069.2+03.8 & K 3-46   & 30  & \nodata & \nodata & \nodata & \nodata &
\nodata  &  n, g4 \\ 
 069.6$-$03.9 & K 3-58   & 19  &    31.4 &      +8 &      61 &    1150
& $-$12 $-$7 & d, g2  \\ 
 074.5+02.1 & NGC 6881 & 23  &     4.9 &   $-$12 &      43 &     150 &
$-$30 $-$15 & d, g3  \\ 
 077.7+03.1 & KjPn 2   & 51  & \nodata & \nodata & \nodata & \nodata &
$-$22 $-$17  &  n, g2  \\ 
 078.9+00.7 & Sd 1     & 69  & \nodata & \nodata & \nodata & \nodata &
$-$40 +0  & n, g1, s  \\ 
 079.6+05.8 & M 4-17   & 34  &     6.2 &   $-$23 &      51 &     170 &
$-$10 $-$6  & d, g4  \\ 
 082.1+07.0 & NGC 6884 & 20  & \nodata & \nodata & \nodata & \nodata &
\nodata  &   n, g3  \\ 
 084.2+01.0 & K 4-55   & 30  & \nodata & \nodata & \nodata & \nodata &
$-$89 $-$53, $-$26 $-$11 &  n, g4 \\ 
 091.6$-$04.8 & K 3-84   & 28  &     6.5 &   $-$80 &      63 &     130
& \nodata  & d, g2  \\ 
 092.1+05.8 & K 3-79   & 29  & \nodata & \nodata & \nodata & \nodata &
$-$23 $-$15  &  n, g3 \\ 
 094.5$-$00.8 & K 3-83   & 33  &    10.8 &   $-$66 &      45 &  420
& $-$63 $-$44, $-$33 $-$27 &  d, g2 \\ *
 103.7+00.4 & M 2-52   & 23  & \nodata & \nodata & \nodata & \nodata &
$-$84 $-$49   & n, g4  \\ 
 104.4$-$01.6 & M 2-53   & 20  &     7.8 &   $-$74 &      55 &     290
& $-$38 $-$48  &  d, g4  \\ 
 107.6$-$13.3 & Vy 2-3  & 17  &    0.40 &   $-$51 &    $>$5 &      80 &
\nodata  & t, g3  \\ 
 107.7$-$02.2 & M 1-80   & 32  & \nodata & \nodata & \nodata & \nodata
& \nodata  &  n, g3  \\ 
 111.2+07.0 & KjPn 6   & 23  & \nodata & \nodata & \nodata & \nodata &
$-$70 $-$66, $-$11 $-$7 & n, g1 \\ 
 119.6$-$06.7 & Hu 1-1   & 60  &     6.8 &   $-$68 &   $>$33 &     190
& \nodata &  d, g3 \\ 
 129.5+04.5 & K 3-91   & 49  & \nodata & \nodata & \nodata & \nodata &
\nodata &   n, g1  \\ 
 130.3$-$11.7 & M 1-1   & 25  & \nodata & \nodata & \nodata & \nodata &
\nodata  &   n, g1  \\ 
 130.4+03.1 & K 3-92   & 33  & 6.5     &   $-$62 &      50 &     180 &
$-$13 $-$10   & d, g3 \\ 
 149.0+04.4 & K 4-47   & 44  & 19.7    &   $-$27 &      49 &     780 &
\nodata &  d, g2  \\ 
 153.7$-$01.4 & K 3-65   & 44  & 9.3     &   $-$30 &      45 &     290
& $-$29 $-$23 &  d, g3  \\ 
 165.5$-$06.5 & K 3-67   & 33  & \nodata & \nodata & \nodata & \nodata
& +0 +4  & n, g3  \\ 
 167.4$-$09.1 & K 3-66   & 27  & \nodata & \nodata & \nodata & \nodata
& +8 +15  &  n, g1  \\ 
 184.6+00.6 & K 3-70   & 24  & 2.7     &     +27 &      35 &      96 &
\nodata &  d, g2 \\ 
 204.8$-$03.5 & K 3-72   & 48  & \nodata & \nodata & \nodata & \nodata
& \nodata  &    n, g1  \\ 
 359.8+06.9 & M 3-37   & 25  & 9.9     &   $-$65 &      45 &     560 &
$-$9 $-$1 &  d, g2  \\ 
\enddata
\tablecomments{ Column [9]: d, detection; n, non-detection; 
t, tentative detection; g1--4, spectroscopic group; s, severe
interstellar contamination}

\end{deluxetable}

\end{document}